\documentclass[a4paper,fleqn,usenatbib,useAMS]{mnras}

\usepackage{graphicx}	
\usepackage{amsmath}	
\usepackage{amssymb}	
\usepackage{multicol}        
\usepackage{bm}		
\usepackage{pdflscape}	
\usepackage{longtable}	
\usepackage{soul}
\usepackage{booktabs}
\usepackage{multirow}
\usepackage{lscape}
\usepackage{pdflscape}
\usepackage{float}

\usepackage[T1]{fontenc}
\usepackage{ae,aecompl}

\usepackage{newtxtext,newtxmath}

\title[Unveiling the internal structure of Hercules supercluster]{Unveiling the internal structure of  Hercules supercluster}

\author[Monteiro-Oliveira et al.] 
  {R.~Monteiro-Oliveira,$^{1,2,3}$\thanks{E-mail: rogerionline@gmail.com}
  D. F.~Morell,$^{4}$
  V. M.~Sampaio,$^5$
  A. L. B.~Ribeiro,$^2$
  and
  \newauthor
  R. R. de~Carvalho$^5$\\
  $^{1}$Academia Sinica, Institute of Astronomy and Astrophysics,
11F of AS/NTU Astronomy-Mathematics Building, No.1, Sec. 4, Roosevelt Rd, Taipei 10617, Taiwan, R.O.C.\\
  $^{2}$Universidade Estadual de Santa Cruz, Laborat\'orio de Astrof\'isica Te\'orica e Observacional - 45650-000, Ilh\'eus-BA, Brazil\\  
  $^{3}$Universidade de S\~ao Paulo, Inst. de Astronomia, Geof\'isica e Ci\^encias Atmosf\'ericas, Depto. de Astronomia, R. do Mat\~ao 1226, 05508-090 S\~ao Paulo, Brazil\\
  $^{4}$Instituto de Astronom\'ia Te\'orica y Experimental (CCT C\'ordoba, CONICET, UNC), Laprida 854, X5000BGR, C\'ordoba, Argentina \\
  $^{5}$NAT - Universidade Cruzeiro do Sul / Universidade Cidade de S\~ao Paulo, 01506-000, Brazil\\ 
  }

\date{Accepted 2021 November 3. Received 2021 November 3; in original form 2021 August 19.}

\pubyear{2018}

\begin{document}
\label{firstpage}
\pagerange{\pageref{firstpage}--\pageref{lastpage}}
\maketitle

\begin{abstract}
We have investigated the structure of the Hercules supercluster (SCL160) based on data originally extracted from the Sloan Digital Sky Survey SDSS-DR7. We have traced the mass distribution in the field through the numerical density-weighted by the $r^\prime$-luminosity of the galaxies and classified them based on their spatial position and redshift. This has allowed us not only to address the kinematics of the supercluster as a whole, but also the internal kinematic of each cluster, which was no further explored before. We have confirmed that the  Hercules supercluster is composed of the galaxy clusters A2147, A2151, and A2152. A2151 consists of five subclusters, A2147 on two and A2152 on at least two. They form the heart of the Hercules supercluster. We also have found two other gravitationally bond clusters, increasing, therefore, the known members of the supercluster. We have estimated a total mass of $2.1\pm0.2 \times 10^{15}$ M$_\odot$ for the Hercules supercluster. To determine the dynamical masses in this work, we have resorted to the $M_{200}-\sigma$ scaling relation and the caustic technique. Comparing both methods with simulated data of bimodal merging clusters, we found the caustic, as well as the $\sigma$-based masses, are biased through the merger age, showing a boost just after the pericentric passage. This is not in line with the principle of the caustic method that affirms it is not depending on the cluster dynamical state.
\end{abstract}




\section{Introduction}
\label{sec:intro}

Structures have been grown in the universe since early times, driven by the gravitational force. The continuous growing has started from initial density perturbations and then from the hierarchical merging of smaller bodies \citep[e.g.][]{Kravtsov12}. At the theoretical view, the process of structure formation can be described in terms of the density contrast parameter \cite[$\delta$; e.g.][]{Guth82}. A key point to understand how the universe has evolved from $\delta\approx 10^{-5}$  in the recombination epoch \citep{planck15b} to the present $\delta\approx 10^2$ as seen in galaxy clusters \citep[e.g.][]{More11} is the presence of the cold dark matter (CDM). Just after the beginning of the equivalence matter-energy epoch, baryons and photons were coupled, prevent any growth of baryon overdensities \citep[the Silk damping;][]{Silk1967,Silk1968}. However, as CDM is not subject to this damping, the first structure seeds started to grow in the density field before the recombination epoch, when the baryons finally started to fall into the pre-existent potential wells created by CDM condensation.

According to the hierarchical scenario, guided by the $\Lambda$CDM cosmology, the superclusters of galaxies constitute the next generation of virialized structures in the Universe, reaching masses $\sim10^{16}$ M$_\odot$ \citep[e.g.][]{Einasto21} and  extending across tens of Mpc \citep[e.g.][]{Bagchi17}. Currently, the top of mass function is occupied by the galaxy clusters with masses $10^{14}$--$10^{15}$ M$_\odot$ \citep[e.g.][]{Kravtsov12}. The number of known superclusters has been increasing \citep[e.g.][]{Einasto97,chon13,Chow-Martinez14}  in spite of some of them were later reclassified since they will not collapse in the future \citep{chon15}, as for example the ex-superclusters Shapley \citep{Scaramella89} and Laniakea \citep{Tully14}.

Superclusters of galaxies can provide us with a varied field of studies. For example, given the diversity of environments, ranging from poorly populated voids to densest regions \citep[e.g.][]{Santiago-Bautista20},  superclusters are a good laboratory to study galaxy evolution \citep[e.g.][]{Ribeiro13,Krause13,Guglielmo18,Seth20,Kelkar20}, which is strongly affected by their environment. The Morphology-Density relation shows that galaxy types are not uniformly distributed in space \citep{Dressler}. Red, quiescent, early-type galaxies are mainly found in high density environments, while blue, star forming, late-type galaxies tend to avoid them. Among different environmental quenching mechanisms such as: Starvation \citep{2017MNRAS.466.3460V}; ram pressure stripping \citep[RPS]{1972ApJ...176....1G}; and tidal mass loss \citep{1999MNRAS.302..771J}, interaction/mergers between clusters can also be relevant to galaxy evolution. The more striking example is the bullet-cluster case, where there is a noticeable enhancement of ram pressure effect, which quickly removes galaxy's gas component (e.g. \citealt{Lourenco20,Moura20}). More recently, it has been considered that cosmic filaments also carry a considerable amount of angular momentum \citep{Wang21}. Other peculiar features can be also found in some objects, as the unusual mass concentration in the core of Saraswati supercluster \citep{Bagchi17}, which can call into question the predictions of modern cosmology.

The Hercules supercluster \citep[SCL160;][]{Einasto97} was first mentioned in the literature by \cite{Shapley1934}, who noticed an galaxy overdensity in the direction of Hercules constellation. Later, \cite{Cooke77} and \cite{Tarenghi79}  stated that it is formed by the galaxy clusters  A2151 \citep[$z=0.0366$, the ``Hercules cluster'';][]{strublerood99}, A2147 \citep[$z=0.0350$;][]{strublerood99} and A2152 \citep[$z=0.0410$;][]{strublerood99}. Following, we will introduce the particularities of each one them. Their location and respective X-ray mass estimation \citep{Piffaretti11} can be found in Table.~\ref{tab:masses}.

\begin{table}
\caption[]{X-ray derived masses according to \cite{Piffaretti11}. The original $M_{500}$ were translated into $M_{200}$ supposing that the mass density can be described by an NFW profile with a halo concentration given by \cite{duffy08}}
\label{tab:masses}
\begin{center}
\begin{tabular}{l c c c c}
\hline
\hline 
 Cluster & $\bar{\alpha}$ & $\bar{\delta}$ & $M_{200}$            & $R_{200}$\\
         &       (J2000)  & (J2000)        & ($10^{14}$ M$_\odot$) & (Mpc)\\
\hline
A2147 & 240.5779 & +16.0200 & 3.53 & 1.44 \\ 
A2152 & 241.3842 & +16.4420 & 0.81 & 0.88 \\ 
A2151 & 241.7179 & +17.7810 & 0.47 & 0.74 \\ 
\hline
\hline
\end{tabular}
\end{center}
\end{table}

The galaxy cluster Abell 2151 (A2151) is located north of the field. Using data from {\it Einstein} Observatory, \cite{Magri88} showed that A2151 has a bimodal X-ray emission. In contrast to the hot gas, \cite{Bird93} identified three kinematic subclusters which later \citep{Bird95} called A2151N, A2151E, and A2151C. However, the X-ray emission is found to be located only in the central substructure (A2151C).

Taking the cluster as a single one, \cite{Escalera94} estimated the virial mass of $M_{\rm vir}=1.07\pm2.13 \times 10^{15}$ M$_\odot$ inside a radius of $1.95\pm0.14$ Mpc, based in radial velocities of only 79 galaxies. More recently, \cite{Agulli16} updated this value to $M_{200}=4.0\pm 0.4\times10^{14}$ M$_\odot$ using the caustic technique and 360 members.

The internal structure of A2151 proved to be more complicated than previously hinted. After a visual inspection on the projected galaxy distribution, \cite{Maccagni95} suggested qualitatively the presence of a fourth structure, A2151S,  located at South.  Unfortunately, they did not perform any statistical procedure to classify the galaxies only drawing straight arbitrary lines as subclusters borders. \cite{LopesDeOliveira10} showed that A2151S is not a fossil group as previous stated. They speculate that it would be part of A2151.

With improved X-ray observations by ROSAT PSPC, \cite{Bird95} turned clear that the subcluster A2151C is composed of two ICM substructures. \cite{Huang96} named them as A2151C-B and A2151C-F, in reference to their relative brightness. Posterior analysis based in ROSAT HRI \citep{Huang96} and XMM-{\it Newton} data showed that  A2151C-B is a cool core, implying it is a dynamically relaxed structure \citep[e.g.][]{Soja18}. Regarding the other subclusters, \cite{Bird95} and \cite{Huang96} agreed that A2151E has a very weak emission whereas A2151N and A2151S present no detectable X-ray emission \citep[e.g.][]{Monteiro-Oliveira20,Doubrawa20}.

The ambiguity in the optical and X-ray characterisation makes A2151 an unsolved puzzle under the kinematic point-of-view. What dynamical scenario led the subclusters to the current configuration? \cite{Bird95} proposed a partial explanation where the eastern and central subclusters are seen in a post-merger phase due to gas absence in the former. However, they disregard the fact that A2151C's gas has a bimodal structure. Furthermore, they did not mention the role of A2151N and  A2151S. Thereby, a full dynamical description involving all subclusters is still lacking in the literature.

In contrast with A2151, the galaxy cluster Abell 2147 (A2147) has not been intensively studied yet. Based on {\it Chandra} data, \cite{Sanderson06},  classified  A2147 as a merger candidate, suggesting that it has many components. They also stated that there is no spatial coincidence between the BCG and the cluster potential well traced by the ICM. \cite{Zhang11} found this offset to be  9.2 kpc. In line with previous findings, \cite{Hudson10} classified A2147 as a non-cool core and a probable merger because of its elongated X-ray emission \citep{Vikhlinin09}. However, its caustic structure does not differ from a single cluster \citep{Wojtak07b}.

Among the clusters, Abell 2152 (A2152) has been the most poorly studied regardless of the wavelength. According to \cite{Blakeslee01}, it forms a ``double cluster'' along with A2147 because of their projected proximity. A2152 contains two BCGs (0.5 mag difference) distant 0.47 arcsec each other \citep{Blakeslee01}. The authors have suggested that the apparent offset between the first BCG and X-ray peak (2.1 arcmin) is due to the misidentification of a background source. But lacking high-resolution data, they were not able to answer this question.

As well as the internal structure of the clusters beforehand mentioned, the Hercules supercluster as a whole has been not much explored so far. The unique attempt to describe this large scale structure was done by \cite{Barmby98} who provided a kinematic view of the supercluster, but they considered the clusters as single ones whereas, at least A2151 is clearly multimodal.

Will have been more cluster members than those already stated? What is the total mass of the supercluster? These are some questions that we aim to fill in the present work. We have conducted an extensive study about the entire field of Hercules supercluster, providing well-detailed anatomy including not only the previously identified members but also new candidates. We have identified the supercluster members not only based on their galaxy projected distribution but also on their radial velocities. To accomplish these tasks, we resorted to the galaxy catalogue of \cite{Yang07}, latter improved by \cite{deCarvalho17}. Having the clusters identified, we have computed their masses based on the galaxy dynamics. To do that, we have compared the power of two well-known methods: the scaling relation $M_{200}-\sigma$ \citep[e.g.][]{Evrard08,Munari13} and the caustic \citep{diaferio97,diaferio99}. Finally, with the masses, we have obtained the kinematic description of each cluster. We also proposed a toy model for the supercluster as a whole.

This paper is organized as follows. Data description is presented in Section \ref{sec:data}. The mapping of the galaxy distribution can be found in Section \ref{sec:photo} followed by the dynamical analysis in Section \ref{sec:dynamical}. The mass estimates of the identified structures are in Section \ref{sec:masses}. The cluster's internal kinematics are described in Section \ref{sec:two.body}. We discuss the results of our analysis in Section \ref{sec:discussion} and summarize our findings in Section \ref{sec:summary}.

In this paper, we adopt the standard $\Lambda$CDM cosmology, given by $\Omega_m = 0.27$, $\Omega_\Lambda = 0.73$, $\Omega_k = 0$, and $h=0.7$.

\section{Data description}
\label{sec:data}

Our sample is build using the Yang Catalog \citep{2007ApJ...671..153Y}, which apply a Halo Finder Algorithm \citep{2005MNRAS.356.1293Y} to the New-York University Value Added Galaxy Catalog \citep{2005AJ....129.2562B}. The original Yang Catalog is based on the Sloan Digital Sky Survey 5$\rm ^{th}$ data release \citep{2004MNRAS.351.1151B}. However, here we use an updated version presented in \cite{2017AJ....154...96D} (dC17, hereafter), which is based on the SDSS-DR7 \citep{2017AJ....154...96D} and we next describe in details.

The dC17 catalog is built by selecting all SDSS-DR7 galaxies within $\rm 0.03 \leq z \leq 0.1$\footnote{This limit refers to the mean redshift of the selected group/cluster. Then, the underlying members were selected in a  slice comprising galaxies within $\pm4000$ km s$^{-1}$. } and with apparent magnitude $\rm m_{r} \leq 17.78$, which is the survey spectroscopic completeness limit at $z=0.1$. The lower redshift bound is adopted to avoid bias in the stellar population parameters estimates due to the fixed 3 arcsec aperture used in the SDSS. Membership is then defined by applying a shiftgapper technique (see \cite{2009MNRAS.399.2201L} for more details) to the galaxies with line-of-sight velocity and projected radial distance of within the range $\rm \pm 4000 \, km \, s^{-1}$ and $\rm d_{proj}\leq 2.5h^{-1}$Mpc (i.e. $\sim$3.47\,Mpc for h = 0.7), respectively, with respect to the clustercentric coordinates (RA, DEC and redshift) described in the Yang Catalog\footnote{We highlight that clustercentric coordinates are the only information we use from the Yang Catalog}. The advantage of using the dC17 catalogue instead of the original SDSS data relies on the fact that the former is more complete, bringing complementary information on the galaxies stellar populations (e.g. age, metallicity, stellar mass, among others). This wealth of information provides subsides to further investigations as, for example, of how these properties can be correlated with the cluster merger phase.

We retrieved the catalogue of all Yang groups located in a circular region with 3 degrees of radius centred in $\alpha, \delta$~=~$(240.57792,+16.020)$. This region is large enough to encompass the three main constituents of Hercules supercluster, A2147, A2151 and  A2152 as well as other possible companion groups.

\begin{figure}
\begin{center}
\includegraphics[width=1.0\columnwidth, angle=0]{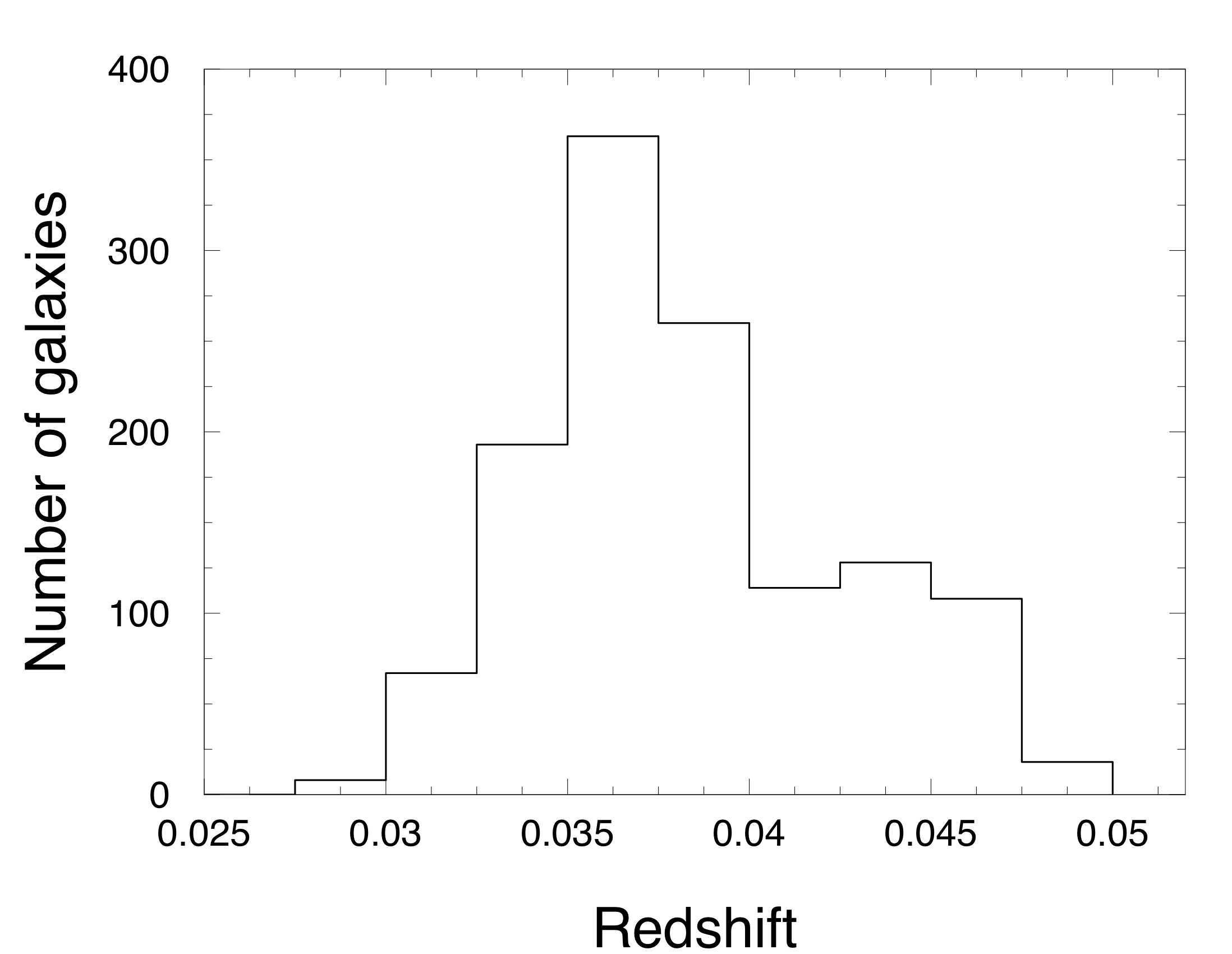}   
\caption{Redshift distribution of the 1259 galaxies in the Hercules catalogue.}
\label{fig:comp}
\end{center}
\end{figure}

The so-called Hercules catalogue is comprised by 1259 galaxies within the interval $8490 \ {\rm km\ s^{-1}}\leq v \leq 14690  \ {\rm km\ s^{-1}}$\footnote{We adopted $v=cz$.} or equivalently $0.02832\leq z\leq 0.04899$. In Fig.~\ref{fig:comp}, we present the respective redshift distribution. The Hercules Cluster is located in the nearby universe, which allows a deeper completeness limit in luminosity, $\rm M_{r_{com}}$. In this work we adopt the limit established in dC17 for structures with $z \leq 0.04$, namely $\rm M_{r} \leq -18.40$.

\section{Mapping the galaxy distribution}
\label{sec:photo}

The cluster's galaxy content is embedded in a large dark matter halo that corresponds to $\sim$80 per cent of the cluster total mass whereas the former accounts only for $\sim$5 per cent. However, despite some bias be present, the integrated stellar light is a good tracer of the cluster total mass.  In this sense, we employed the projected galaxy distribution to map the mass allocation along the Hercules field, using the sample selected in Section~\ref{sec:data}.

To translate the discrete galaxies into a smoothed map \citep[e.g.][]{Wen13}, we split the field into squared cells of 1 arcmin$^2$ inside which we computed the numerical density
\begin{equation}
    D(\vec{\xi}) = \sum_{i=1}^N  K(\vec{\xi_i},\sigma_\xi)L_i,
\label{eq:dens}
\end{equation}
representing a sum over all $N$ galaxies located inside a radius of $\sigma_\xi$, the smoothing scale. In case we want to weight the map by the  $r^\prime$-luminosity, we adopt
\begin{equation}
L_{i}=10^{-0.4(M_{r_i}-M_{r_{\rm com}})}
\label{eq:luminosity}
\end{equation}
whereas $L_i=1$ if we want to compute only the single numerical density. Then, $L_i$ is convoluted by the Epanechnikov kernel,
\begin{eqnarray}
K(\vec{\xi_{i}},\sigma_\xi) = \left \{
\begin{array}{ll}
\frac{3}{4} \left [1-\left(\frac{\vec{\xi_{i}}}{\sigma_\xi}\right)^2 \right ], & \vec{\xi_{i}} \le \sigma_\xi \\
0, & \vec{\xi_{i}} > \sigma_\xi.
\end{array}
\right.
\label{eq:kernel}
\end{eqnarray}

\begin{figure*}
\begin{center}
\includegraphics[width=1.0\textwidth, angle=0]{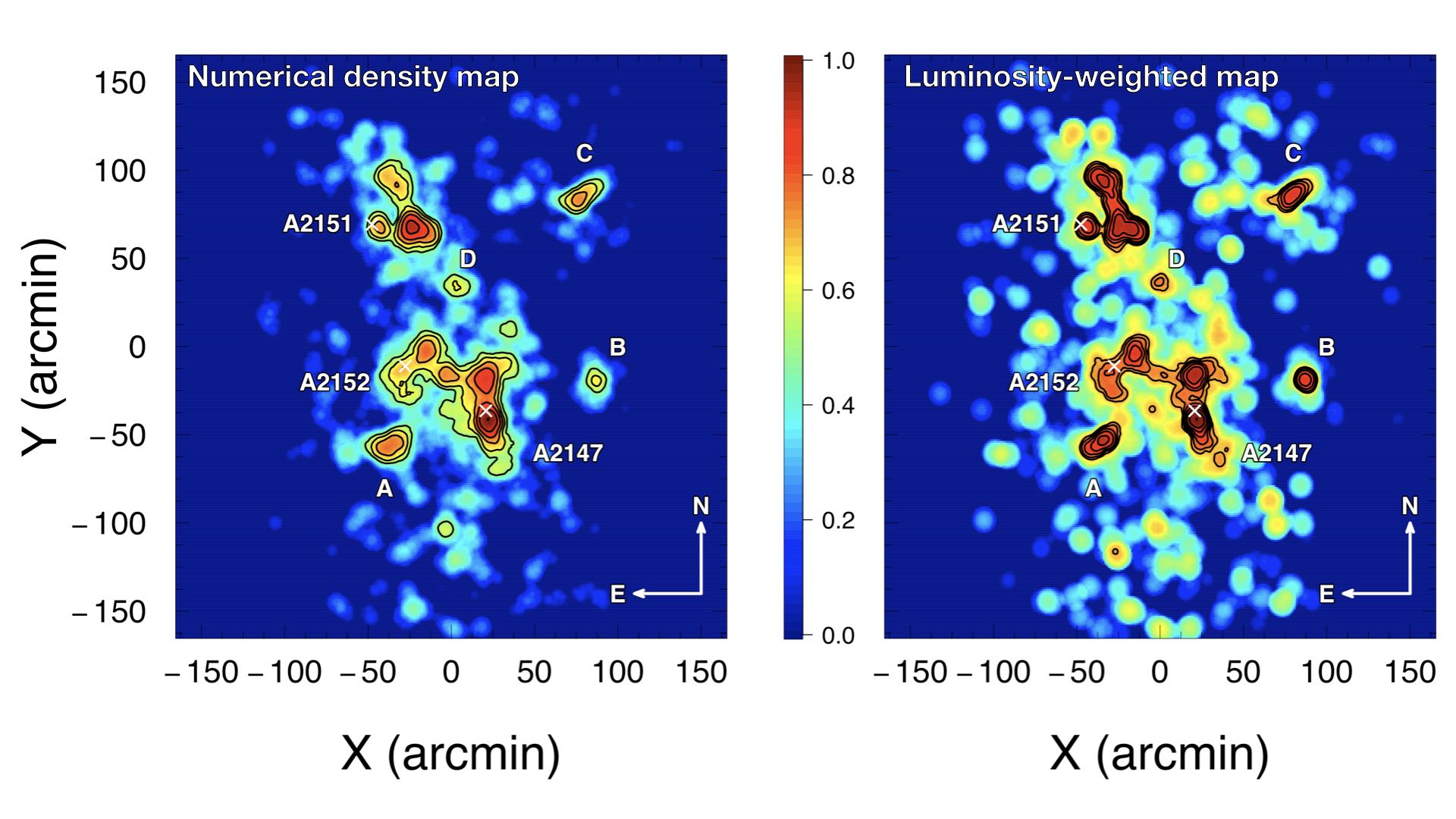}     
\caption{Galaxy distribution in the Hercules supercluster. All values are normalized by the maximum pixel in the respective map. {\it Left: } Numerical projected density map. {\it Right: } Numerical density-weighted by $r^\prime$ luminosity. In both maps, the crosses represent the cluster location according to the X-ray catalogue of \protect\cite{Piffaretti11}. Other four noteworthy clumps are labelled A--D.}
\label{fig:maps}
\end{center}
\end{figure*}

We computed both the numerical density and the luminosity-weighted map. The results can be seen in Fig.~\ref{fig:maps}, where a smoothing scale of 9 arcmin was adopted. Importantly, the final map does not change significantly if an alternative scale within a few arcmin is adopted instead.

Overall, the two maps are very similar to each other, presenting roughly the same complex scenario. By definition, the numerical density map highlights the density contrast, emphasizing the borders. On the other side, the luminosity weighted map is more efficient in enhancing the substructures. As this is a piece of vital information for our purposes, we choose to use this map henceforth. Despite the majority of galaxies appear to be related to the clusters A2147, A2151 and A2152, they present themselves an intricate multimodal structure. Whereas A2151 is relatively isolated at the North, A2147 and A2152 are close to each other being connected through a bridge. The main clusters are surrounded by other four galaxy clumps (A--D).

\begin{figure*}
\begin{center}
\includegraphics[width=1.0\textwidth, angle=0]{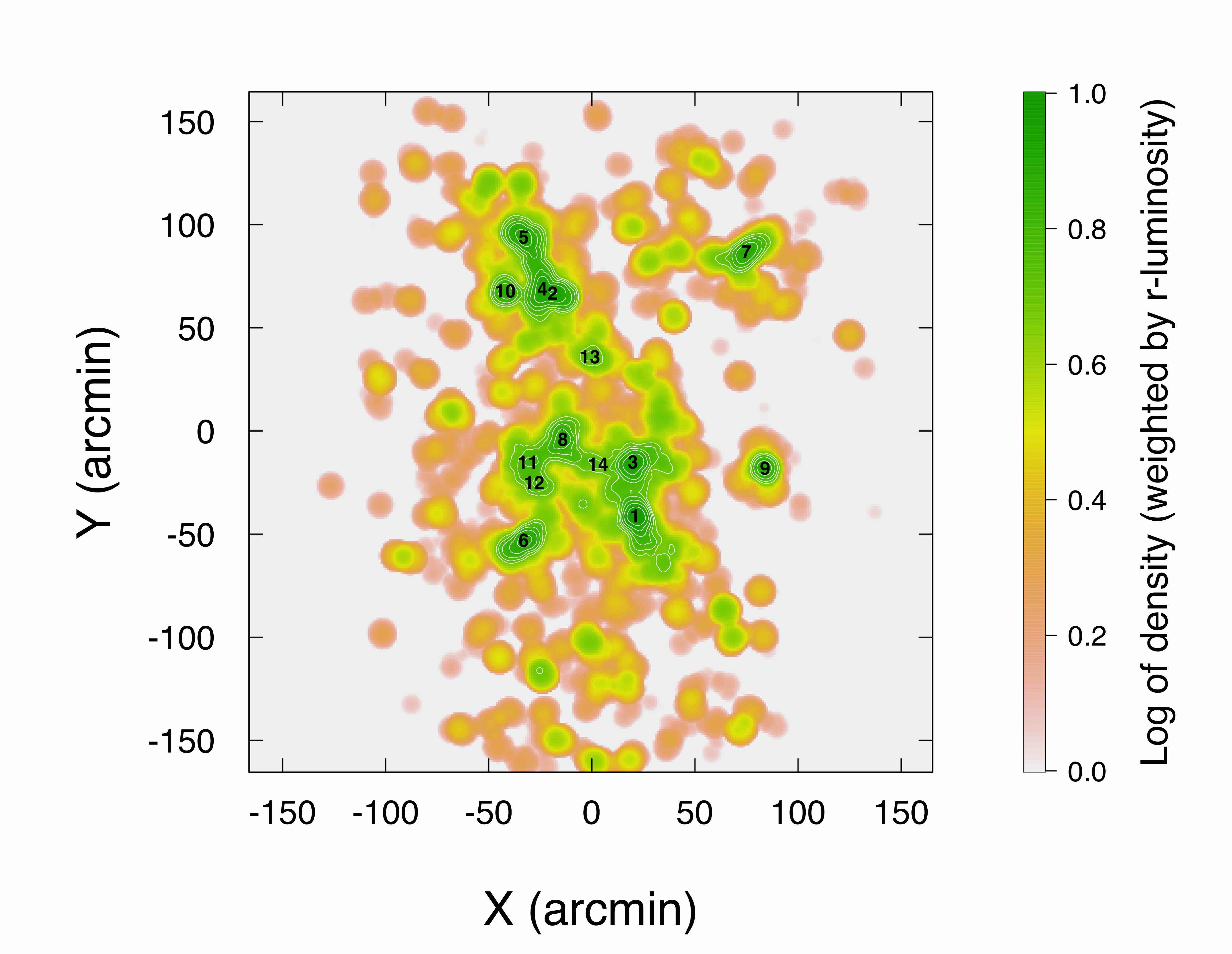}     
\caption{Mass peaks identified in the luminosity-weighted density map above the threshold $10\sigma_n$. The number shows the ranked significance and are placed exactly at the corresponding peak centre. White contours are the same as Fig.~\ref{fig:maps}-right.}
\label{fig:peaks}
\end{center}
\end{figure*}

To identify the most prominent galaxy clumps and therefore the most massive regions, we have employed a tailor-made algorithm. It works searching for the local maxima within a moving circular of a 3 arcmin radius. Then, the peak centre position is defined as the pixel-weighted mean inside this circular region. The uncertainty in each peak position was set as 50 kpc. To determine the noise level, we resampled 10,000 times the map (allowing repetition) taking the standard deviation after each iteration, and the overall mean $\sigma_n$ at the end. The significance on each peak detection, $S$, was defined as the ratio between the local maxima value and the noise level $\sigma_n$. After this procedure, we found 14 high-density regions above the threshold $S>10\sigma_n$ as we can see in Fig.~\ref{fig:peaks}.

There are five meaningful peaks in the area of A2151, encompassing the four known members (\#2, \#4, \#5, and \#10) plus a candidate (\#13). In A2147, there are two structures (\#1 and \#3), and three others are found in A2152 (\#8, \#11, and \#12). In the same region, a peak is located at the bridge (\#14) and surrounding A2152 (\#6). As previous stated, other two satellites are also found (\#7 and \#9).

After to find the supercluster's backbone, our forthcoming analysis aims to investigate the role of these mass clumps in the internal dynamics of each galaxy cluster as well as to check if the surrounding clumps A--D are bounded to the beforehand mentioned clusters or they constitute themselves independent structures.


\section{Dynamical analysis}
\label{sec:dynamical}

Despite the large area covered ($\sim$28 deg$^2$), we will focus on the densest part of the field where the virialized regions are expected to be. With this approach, we do not want to take into account sparse galaxies whose unique contribution is to add noise in the galaxy cluster membership assignment,  distracting the reader from the goal of describing the main clusters substructures, their internal dynamics and large-scale kinematics.

We have selected all galaxies inside a radius of 1.1 Mpc ($\sim$ 24 arcmin) from each clump introduced in Fig.~\ref{fig:peaks}. A visual inspection ensured that any spatial bias was not added, which would potentially lead to the detection of a non-physical structure in the field. For the sake of organization, we will refer to three regions hereafter: 1) A2151 + D, 2) A2152 + A2147 + A and 3) B + C. To avoid contamination by galaxies belonging to any neighbour structure, the 3$\sigma_v$-clipping procedure \citep{3sigmaclip} was applied in each region to remove outliers.

To proceed the galaxy membership assignment inside the regions, we resort to the Gaussian multidimensional mixture modelling {\sc Mclust} \citep{mclust}, implemented in the {\sc R} package \citep{R}. In general lines,  the algorithm searches for optimized clusters from models encompassing variable shapes, orientations and volumes. For a detailed description of the application of {\sc Mclust} in galaxy classification, we refer the reader to recent works of \cite{Morell20} and \cite{Lourenco20}.

\begin{figure}
\begin{center}
\includegraphics[width=1.0\columnwidth, angle=0]{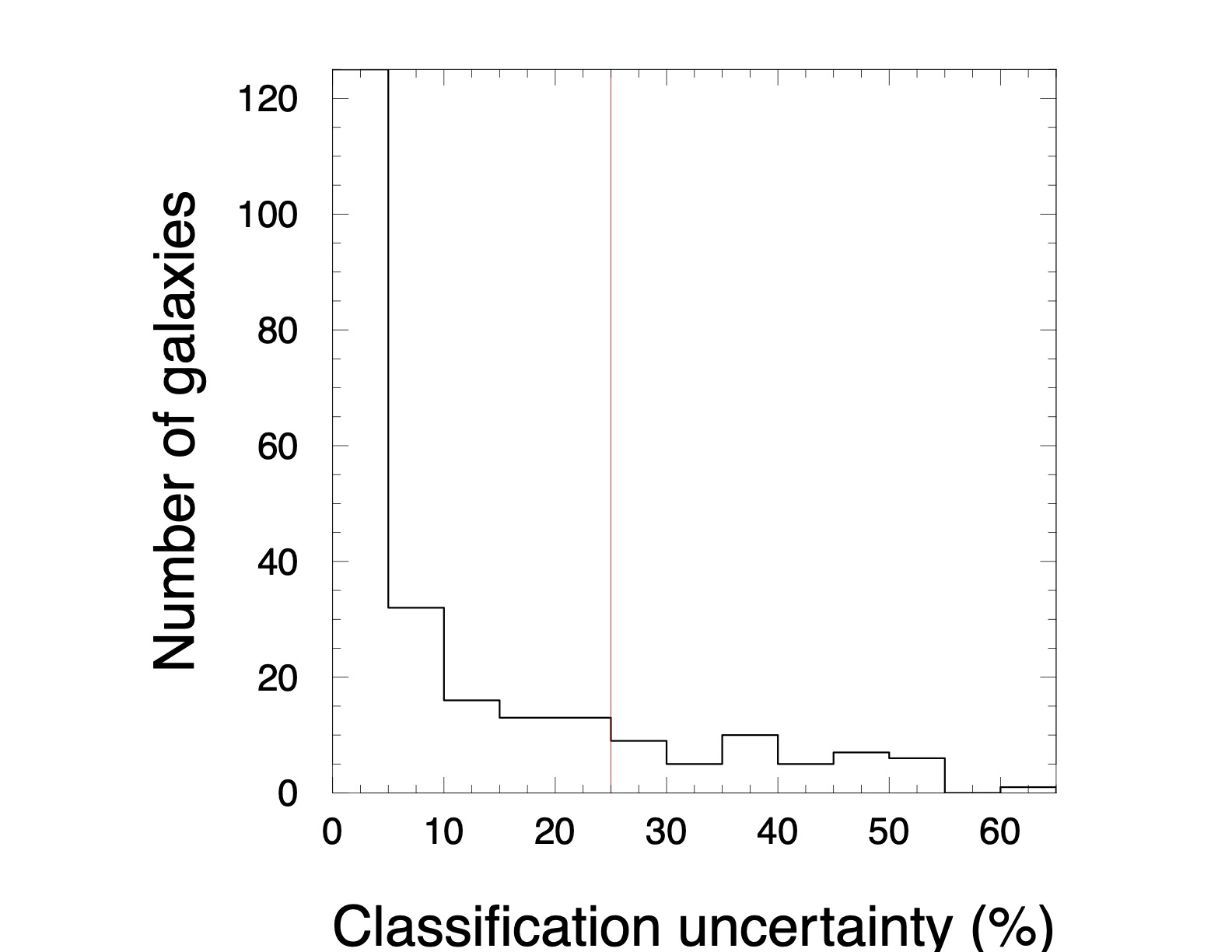}     
\caption{Uncertainty in the galaxy classification given by  {\sc Mclust} for the region A2151 + D. In order to increase the confidence in the subclusters' membership assignment, our final sample only contains those galaxies whose classification was done with an uncertainty of less than 25 per cent.}
\label{fig:class.u}
\end{center}
\end{figure}

We have applied the {\sc Mclust} in its 3-dimensional mode, having as input the spatial coordinates plus the radial velocity of each galaxy. Despite being allowed by the {\sc Mclust}, no informative prior was given. The most credible model for galaxy classification, i.e. the number of groups and the corresponding galaxy membership, was chosen after the Bayesian Information Criterion \cite[BIC;][]{kass95}\footnote{It is very important to stress that {\sc Mclust } implements an opposite definition of BIC than largely found in the literature \citep[e.g.][]{schwarz78}. In the particular case of {\sc Mclust } a larger BIC point in favour of the preferred model.}. However, the classification of each galaxy is not a unique quantity and there is an uncertainty attributed by the {\sc Mclust}, as we can see in Fig.~\ref{fig:class.u}. To define if a given galaxy is a member of some group, we have considered a maximum of 25 per cent of uncertainty in its classification. This choice has been proved to have a good compromise between the final number of galaxies in each group and the consistency of the dynamical mass obtained from, as we will describe better in Sec.~\ref{sec:masses}.

\begin{figure*}
\begin{center}
\includegraphics[width=1.0\textwidth, angle=0]{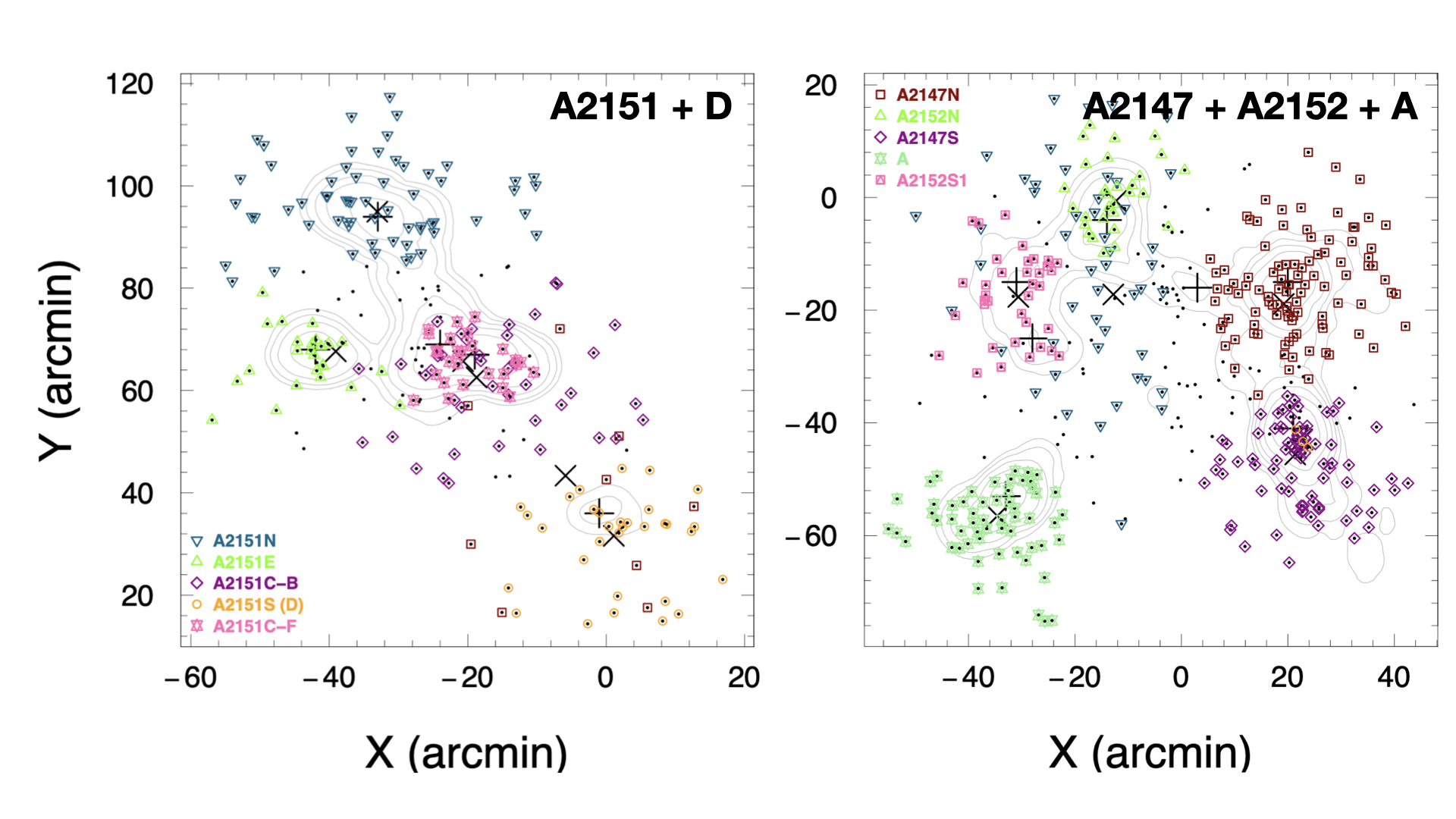}     
\caption{{\sc 3D-Mclust} classification. Black dots alone (i.e. not overlaid with any other symbol) correspond to galaxies whose uncertainty in the classification is greater than 25 per cent, and therefore, were not considered for dynamical analysis. {\it Left: } A2151 + D.  {\it Right: } A2147 + A2152 + A. The same coordinate convention of Fig.~\ref{fig:peaks} was adopted. Plus signals mark the overdensity peaks as identified in Fig.~\ref{fig:peaks} and the crosses identify the groups centroid as calculated by {\sc 3D-Mclust}.}
\label{fig:Mclust}
\end{center}
\end{figure*}

 The final classification is shown in Fig.~\ref{fig:Mclust}. {\sc 3D-Mclust} confirms the complexity of the Hercules field. Each region is itself formed by multiples groups, which we will refer henceforth as subclusters. Noteworthy, the majority of galaxy clumps identified in Sec.~\ref{sec:photo} are also related to dynamical structures. Following, we will discuss the results for each region individually. 

\subsection{A2151 + D}
\label{sec:dynamical.2151}

The galaxy cluster A2151 is comprised by (at least) three subclusters \citep[e.g.][]{Bird93,Bird95,Maccagni95}, previously called A2151N, A2151E, A2151C, according to their position. Later,  \cite{Huang96} found that the central one is not a single, but a bimodal system comprised by A2151C-B and A2151C-F\footnote{B and F stands for bright and faint, respectively and are related to their X-ray emission.}. A fifth structure (labelled ``D'' in Fig.~\ref{fig:maps}), was called A2151S by \cite{Huang96} and \cite{Sanchez-Janssen05} but the authors have not stated any strong argument to ensure that clump is indeed part of A2151. Later, we will provide a definitive proof about this matter after check if it is bounded with A2151 main bodies.

The 199 selected galaxies are characterized by  $\bar{z}=0.0364\pm 0.0027$ and $\sigma_v/(1+\bar{z})=794_{-37}^{+43}$ km s$^{-1}$. According to the Anderson-Darling test, the radial velocity distribution can be described by a Gaussian  function within 95 per cent c.l. ({\it p}-value = 0.10). 

In its best model, {\sc 3D-Mclust} has identified six groups. The second-best model ($\Delta {\rm BIC}=1$\footnote{According to \cite{kass95}, the criteria model for selection from $\Delta{\rm BIC}={\rm BIC}_{\rm major} - {\rm BIC}_{\rm minor}$ are follows: $\Delta {\rm BIC}=1-2$: the models are comparable, $\Delta {\rm BIC}=2-6$: positive evidence in favour of the model with largest BIC, $\Delta {\rm BIC}=6-10$: strong evidence, $\Delta {\rm BIC}>10$: very strong evidence.}) also pointed into six sets with minimum changes concerning the first model. Five of those groups are straightforwardly correlated with the mass overdensities identified in Sec.~\ref{sec:photo} (crosses in Fig.~\ref{fig:Mclust} indicate the peak positions). A full description of the subclusters' properties so far can be seen in Table~\ref{tab:A2151}. They follow, within 95 per cent c.l. ({\it p}-value > 0.10), a Gaussian distribution.

The remaining group  have only seven galaxies that do not follow a Gaussian distribution ({\it p}-value = 0.018) within 95 per cent c. l. They are more spread out in the field than the others, sometimes overlapping with groups related to A2151S and A2151C-B. Two reasons make us believe that these galaxies are probably infalling into the A2151's core: the absence of a corresponding mass clump and the non-Gaussianity of their members. So, we did not consider these galaxies for forthcoming analysis.

\subsection{A2147 + A2152 + A}
\label{sec:dynamical.2152}

The Southern part of Hercules supercluster has been poorly studied in optical so far. This field is dominated by the galaxy clusters A2147 and A2152, with a bright SZ emitting bridge connecting them \citep{Planck13}. Despite some earlier studies of Hercules supercluster \cite[e.g.][]{Barmby98,Blakeslee01} have considered those clusters as single structures, the photometric analysis of \cite{flin06} has suggested that A2147 and A2152 are bimodal, which is in reasonable agreement with our finds in Section~\ref{sec:photo}. We have found no mention to the structure ``A'' in the literature.

The {\sc 3D-Mclust} points into an unquestionable scenario with seven dynamical components. The reason for this statement is that the mentioned model is strongly favoured to the second-best model (five groups), as settled by the high $\Delta {\rm BIC}=16$. Notably, most of the galaxies located in the bridge, were not classified according to our selection criteria.

From the seven mass clumps, two of them were not correlated to any dynamical group: the clump \#14 at the bridge and \#12, in the southernmost part of A2152. On the other side, two dynamic groups are not straightforwardly linked with any mass clump. However, one of them has only three galaxies and therefore was disregarded. The second one has 49 galaxies nearly overlaid with A2152 but also spread along the region between the two clusters. This group is characterized by $\bar{z}=0.0451\pm0.0012$, $\sigma_v/(1+\bar{z})=375\pm25$ km s$^{-1}$, and, with 95 per cent c.l., the radial velocities follow a Gaussian ({\it p}-value = 0.13). Its centroid is located midway A2152N and the mass clump \#14. The lack of evidence prevents us to propose a unique explanation for its nature but it is probable comprised of a mixture of galaxies from the bridge and A2152.

The dynamical description of the clusters A2147 and A2152 as well as the mass clump A can be found in Table~\ref{tab:A2147.2152}. We have named the southern fully characterized substructure of A2152 as A2152S1. All of them have their radial distributions described by a Gaussian with 95 per cent c. l. ({\it p}-value > 0.22). Beyond providing us with a way to describe the internal kinematics of the clusters, these results will allow us to determine at what level the mass clump A is or not related to the main body of the Hercules supercluster.

\subsection{B + C}
\label{sec:BC}

The last two structures of the Hercules field are more apart from the  supercluster's core (A2147 + A2151 + A2152). Both B and C, have their members' radial velocity distribution  following a Gaussian distribution within 95 per cent c.l., with p-values being respectively 0.31 and 0.50. More details regarding the subclusters' dynamics can be found in Table~\ref{tab:BC}.

\section{Mass estimates}
\label{sec:masses}

\subsection{$M_{200}-\sigma$ scaling relation}
\label{sec:munari}

The first approach we have applied to estimate the subcluster masses is based on the scaling relation between $\sigma$, the velocity dispersion of their member galaxies, and the host halo mass $M_{200}$. We have adopted the well-known scaling relation  \citep[][]{biviano06, Evrard08,Munari13}:
\begin{equation}
\frac{\sigma}{\rm km \ s^{-1}} = A_{\rm 1D}\left[  \frac{h(z)\ M_{200}}{10^{15} \ {\rm M}_\odot} \right]^\alpha
\label{eq:M.sigma}    
\end{equation}
and we have considering $\sigma=\sigma_v/(1+\bar{z})$,  $A_{\rm 1D}=1177\pm4.2$ km s$^{-1}$ and $\alpha=0.364\pm0.0021$, in agreement with  \cite{Munari13}.

However, such kind of scaling relation is derived from simulated data, and therefore some bias could be added into $M_{200}$. Additionally, some bias contribution could come from observations restrictions as the small number of galaxies and the truncated cluster radius observed. In \cite{Ferragamo20} the authors suggested a set of corrections to improve the mass estimation given by Equation~\ref{eq:M.sigma}. The procedure  starts by correcting $\sigma$ for the small galaxy samples,
\begin{equation}
    \sigma^\prime=\sigma(N_{\rm gal}) \left\{ 1+ \left[\left(\frac{D}{N_{\rm gal}-1}\right)^\beta + B \right] \right \} 
\label{eq:sigma.umb}    
\end{equation}
where $\sigma^\prime$ is the non-biased estimator of the velocity dispersion, $N_{\rm gal}$ is the number of galaxies,  $D=1/4$, $B=-0.0016\pm0.0005$ and $\beta=1$. 

In the next step, $\sigma^\prime$ should be corrected by multiplicative factors that account for the aperture radius where $\sigma(N_{\rm gal})$ is measured ($f_1$), the fraction of massive galaxies present in the sample ($f_2$) and the contamination by interlopers ($f_3$). Since we do not have any information a priory about $R_{200}$, we choose $f_1=0.998\pm0.001$, that corresponds to members enclosed withing $R_{200}$\footnote{Actually, this factor ranges in a small interval [0.973,1.044]. The uncertainty, however, presents a larger spread, as expect, ranging from 0.001 at $1R_{200}$ to 0.128 at $0.2R_{200}$.}, $f_2=0.99\pm0.01$ corresponding to a fraction 50\%--100\% of massive galaxies and $f_3=1.05\pm0.01$ supposing that the sample is contaminated by $\sim5\%$ of interlopers \citep{Wojtak07a}. 

Finally, the estimated mass should also be corrected itself by the effect of the finite sample,
\begin{equation}
M^\prime_{200}=M_{200}(\sigma^{\prime\prime}) \left [ \frac{1-E^\prime \alpha}{(E^\prime \alpha)^2(N_{\rm gal}-1)^{\gamma^\prime}}+F^\prime\right]^{-1}
\label{eq:mass.umb}    
\end{equation}
where $M_{200}(\sigma^{\prime\prime})$ is the biased estimator of the virial mass (Equation~\ref{eq:M.sigma}) , $E^\prime=1.53\pm0.03$, $F^\prime=1$, and $\gamma^\prime=1.11\pm0.04$.

\subsection{Caustic}
\label{sec:caustic}

The second approach to estimate the subcluster masses is the caustic technique developed by \cite{diaferio97} and \cite{diaferio99}. This method is particularly relevant as it provides a reliable way to measure the mass profile in galaxy groups and clusters, making no assumptions about the dynamical state, and just providing the galaxy celestial coordinates and redshifts. For example, \cite{serra11} showed that the caustic technique can recover the mass profile with better than 10 per cent accuracy in the range (0.6--4)$R_{200}$. Furthermore, it provides a way of interloper removal \citep[e.g.,][]{serra10} and the identification of the cluster substructures \citep[e.g.,][]{yu15}.

Assuming a spherically symmetric system, the escape velocity can be related to the potential $\phi$ as
$$ v^2_\text{esc}=-2\phi(r). $$
As we can only measure the line-of-sight (l.o.s) velocity component we have $ \langle v^2_\text{esc}\rangle=\langle v^2_\text{esc,los}\rangle g(\beta) $, where $\beta$ is the anisotropy parameter and
$$ g(\beta)=\frac{3-2\beta(r)}{1-2\beta(r)}. $$
Therefore, the cumulative total mass is
$$ GM(<r)=r^2\frac{d\phi}{dr}=-\frac{r}{2}\langle v^2_\text{esc,los}\rangle g(\beta)\left(\frac{d\ln\langle v^2_\text{esc,los}\rangle}{d\ln r}+\frac{d\ln g}{d\ln r}\right). $$
The equation above poses two problems, first, we must know $ \beta(r) $ which is not generally the case, and second, the measure of $\langle v^2_\text{esc,los}\rangle$ can be noisy by the presence of background and foreground galaxies. As found by \cite{diaferio97}, the last can be bypassed by measuring the amplitude $A(r)$ of the caustics \citep{kaiser87,regos89} in the cluster projected phase space (PPS), representing the average component along the l.o.s. of the escape velocity. The mass profile is then
$$ GM(<r)=\int_0^r A^2(r)F_\beta(r)dr, $$
where $F_\beta(r)=F(r)g(\beta)$, and $F(r)=-2\pi G\rho(r)r^2/\phi(r)$. \cite{diaferio97} also found that, in hierarchical clustering scenarios, $F(r)$ is not a strong function of $r$. Furthermore, $F_\beta(r)$ is also a slowly changing function of $r$, and can be therefore taken as a constant in the equation above:
$$ GM(<r)=F_\beta\int_0^r A^2(r)dr. $$

To locate the caustic surfaces, one must apply a kernel density estimation to the tracers on the PPS of the projected radii and velocities. \cite{diaferio99} uses an adaptive Gaussian kernel method, whereas \cite{gifford13} use a standard fixed multi-dimensional Gaussian kernel that independently adapt to the sampling, according to \cite{Silverman86}, and show that it recovers the cluster mass estimates with low scatter and bias. Then, it can be determined the threshold $\kappa$ that defines the caustic location, chosen by minimizing the quantity $ |\langle v^2_\text{esc,los}\rangle-4\langle v_\text{los}^2\rangle |^2 $ inside $R_{200}$.

Some authors claimed $F_\beta$ to be in the range 0.5--0.7 \citep{diaferio99,serra11,gifford13} against numerical simulations. The assumption of a constant value can leads the caustic technique to overestimates the mass up to 70 per cent at smaller radii \citep{serra10}. We follow the \cite{gifford13} recipe and use $F_\beta=0.65$ and their proposed Gaussian kernel, through an implementation in the R statistical software. As found by \cite{diaferio99}, the most relevant systematic errors in the caustic technique are due to projection effects.

\subsection{Results}
\label{sec:mass.res}

\begin{figure*}
\begin{center}
\includegraphics[width=0.65\textwidth, angle=0]{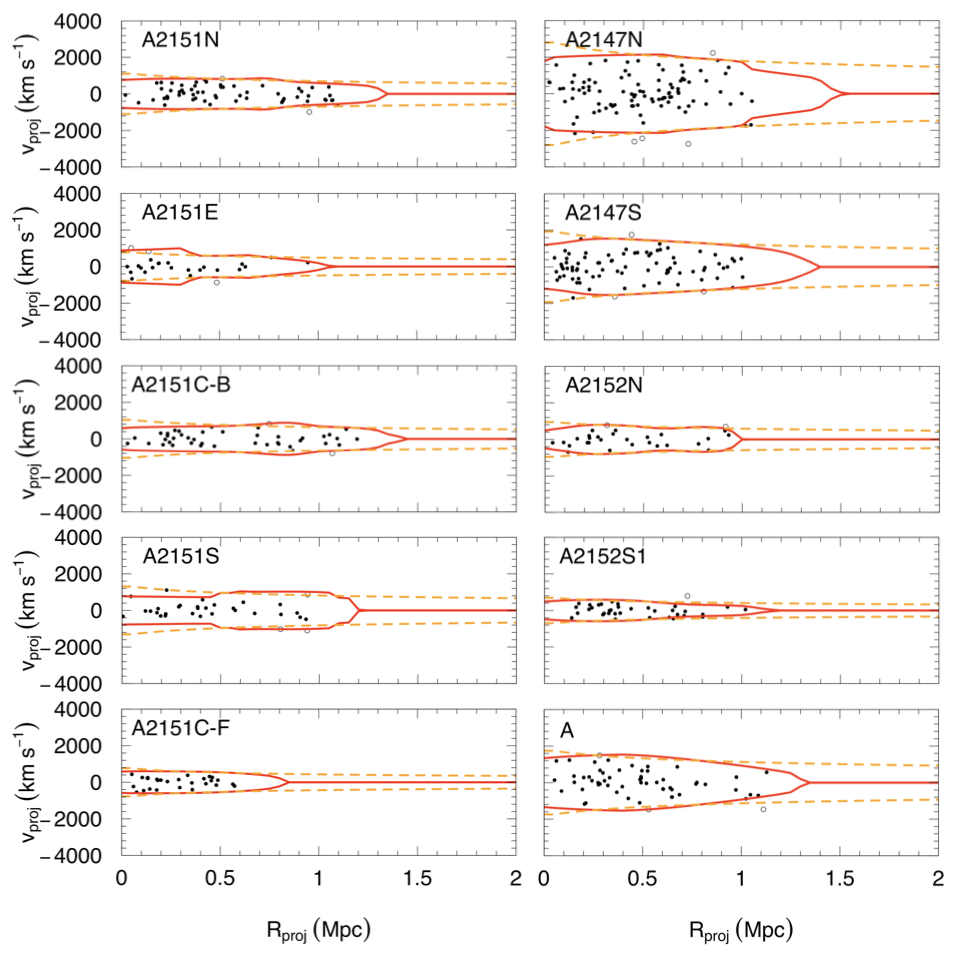}     
\caption{PPS of the groups identified through the 3D-Mclust algorithm. Each plot contains only the  galaxies assigned to their respective host group, as presented in Fig.~\ref{fig:Mclust}. The projected velocity $v_{\rm proj}$ is computed to each i-th galaxy as $v_i-\bar{v}$ in the line-of-sight. The projected distance, $R_{\rm proj}$, is measured in relation to the respective mass peak, as presented in Fig. 3. The continuous red curve corresponds to the caustic profile and the dotted orange curve is the best NFW profile fit to the former and is a representation of the corresponding escape velocity surface. Filled points are considered as members of the groups, whereas open points are considered as interlopers.}
\label{fig:caustic}
\end{center}
\end{figure*}

The computed masses are presented in Table~\ref{tab:A2151}, for A2151 + D, in Table~\ref{tab:A2147.2152}, for A2147 + A2152 + A and in Table~\ref{tab:BC} for B + C. To estimate the errors in the caustic method, we have drawn 1,000 resamplings of the subclusters' centre position and then computed the mass. Each new centre was chosen from a Gaussian distribution with a mean equal to the known position and standard deviation equivalent of its uncertainty (50 kpc). The PPS and the underlying caustic curves are illustrated in Fig.~\ref{fig:caustic}.

\begin{table*}
\caption[]{Substructures of A2151. $\alpha$ and $\delta$ refers to the peaks identified in the luminosity weighted map.}
\label{tab:A2151}
\begin{center}
\begin{tabular}{l c c c c c c c c c }
\toprule

&	 	&   	&	& &  & \multicolumn{2}{c|}{$M_{200}-\sigma$} &\multicolumn{2}{c|}{Caustic}\\ \cmidrule{7-10}
        	
Subcluster	&	$\alpha$	&   $\delta$	&	$N_{\rm gals}$ & $\bar{z}$	&	$\sigma_v/(1+\bar{z})$	&	$M_{200}$	&	$R_{200}$	&		$M_{200}$	&	$R_{200}$	\\[5pt]
        	&	(J2000)  	&   (J2000)  	&	        	&		        &	(km s$^{-1}$)	&	($10^{14}$ M$_\odot$)	&	(kpc)	        & ($10^{13}$ M$_\odot$)	    & (kpc)	\\

\midrule

A2151N 	 & 	 241.49354  	 & 	  18.19872  	 & 	  62  	 & 	  $  0.0374\pm   0.0013$  	 & 	  $404_{- 22}^{+ 25}$  	 & 	  $0.62_{-0.11}^{+0.12}$  	 & 	   $ 809_{-  51}^{+  48}$  	 & 	    $0.95_{-0.30}^{+0.19}‡$   	 & 	     $964_{-94}^{+79}$ \\[5pt]
A2151E 	 & 	 241.65275  	 & 	  17.76490  	 & 	  25  	 & 	  $  0.0391\pm   0.0014$  	 & 	  $415_{- 25}^{+ 24}$  	 & 	  $0.65_{-0.12}^{+0.12}$  	 & 	   $ 823_{-  56}^{+  48}$  	 & 	    $0.69_{-0.36}^{+0.27}$   	 & 	     $871_{-147}^{+124}$ \\[5pt]
A2151C-B 	 & 	 241.24956  	 & 	  17.74695  	 & 	  43  	 & 	  $  0.0363\pm   0.0012$  	 & 	  $358_{- 25}^{+ 20}$  	 & 	  $0.43_{-0.10}^{+0.09}$  	 & 	   $ 716_{-  60}^{+  45}$  	 & 	    $0.68_{-0.15}^{+0.30}$   	 & 	     $864_{-150}^{+63}$ \\[5pt]
A2151S (D)	 & 	 240.93730  	 & 	  17.22772  	 & 	  32  	 & 	  $  0.0342\pm   0.0016$  	 & 	  $471_{- 27}^{+ 29}$  	 & 	  $0.97_{-0.17}^{+0.19}$  	 & 	   $ 938_{-  59}^{+  58}$  	 & 	    $1.13_{-0.41}^{+0.47}$   	 & 	    $1021_{-117}^{+183}$ \\[5pt]
A2151C-F 	 & 	 241.33708  	 & 	  17.78069  	 & 	  28  	 & 	  $  0.0327\pm   0.0009$  	 & 	  $280_{- 18}^{+ 18}$  	 & 	  $0.20_{-0.04}^{+0.04}$  	 & 	   $ 558_{-  38}^{+  38}$  	 & 	    $0.34_{-0.13}^{+0.04}$   	 & 	     $683_{-87}^{+37}$ \\[5pt]
\bottomrule

\end{tabular}
\end{center}
\end{table*}

\begin{table*}
\caption[]{Substructures of A2147 and A2152. $\alpha$ and $\delta$ refers to the peaks identified in the luminosity weighted map.}
\label{tab:A2147.2152}
\begin{center}
\begin{tabular}{l c c c c c c c c c c}
\toprule

&	 	&   	&	& &  & \multicolumn{2}{c|}{$M_{200}-\sigma$} &\multicolumn{2}{c|}{Caustic}\\ \cmidrule{7-10}
        	
Subcluster	&	$\alpha$	&   $\delta$	&	$N_{\rm gals}$ & $\bar{z}$	&	$\sigma_v/(1+\bar{z})$	&	$M_{200}$	&	$R_{200}$	&		$M_{200}$	&	$R_{200}$	\\[5pt]
        	&	(J2000)  	&   (J2000)  	&	        	&		        &	(km s$^{-1}$)	&	($10^{14}$ M$_\odot$)	&	(kpc)	        & ($10^{13}$ M$_\odot$)	    & (kpc)	\\

\midrule

A2147N 	 & 	 240.57737  	 & 	  16.37394  	 & 	  100  	 & 	  $  0.0375\pm   0.0034$  	 & 	  $1021_{- 76}^{+ 60}$  	 & 	  $10.10_{-2.01}^{+1.89}$  	 & 	   $2048_{- 146}^{+ 121}$  	 & 	   $7.66_{-1.32}^{+0.17}$  	 & 	   $1937_{- 118}^{+  13}$ \\[5pt]
A2147S 	 & 	 240.56362  	 & 	  15.93999  	 & 	  81  	 & 	  $  0.0353\pm   0.0024$  	 & 	  $713_{- 47}^{+ 42}$  	 & 	  $3.43_{-0.69}^{+0.72}$  	 & 	   $1430_{- 103}^{+  94}$  	 & 	    $3.32_{-0.35}^{+0.88}$  	 & 	   $1463_{-  53}^{+ 119}$ \\[5pt]
A2152N 	 & 	 241.16792  	 & 	  16.56102  	 & 	  29  	 & 	  $  0.0451\pm   0.0012$  	 & 	  $375_{- 22}^{+ 21}$  	 & 	  $0.48_{-0.09}^{+0.10}$  	 & 	   $ 743_{-  51}^{+  50}$  	 & 	   $0.60_{-0.20}^{+0.25}$  	 & 	   $ 832_{-  56}^{+ 141}$ \\[5pt]
A2152S1 	 & 	 241.46462  	 & 	  16.37852  	 & 	  36  	 & 	  $  0.0441\pm   0.0010$  	 & 	  $294_{- 18}^{+ 18}$  	 & 	  $0.24_{-0.04}^{+0.05}$  	 & 	   $ 585_{-  39}^{+  39}$  	 & 	   $0.28_{-0.10}^{+0.18}$  	 & 	   $ 646_{-  33}^{+ 149}$ \\[5pt]
A 	 & 	 241.50144  	 & 	  15.74458  	 & 	  58  	 & 	  $  0.0400\pm   0.0023$  	 & 	  $690_{- 43}^{+ 41}$  	 & 	  $3.10_{-0.55}^{+0.59}$  	 & 	   $1381_{-  87}^{+  83}$  	 & 	   $2.92_{-0.41}^{+0.81}$  	 & 	   $1407_{-  69}^{+ 120}$ \\[5pt]

\bottomrule

\end{tabular}
\end{center}
\end{table*}

\begin{table*}
\caption[]{Substructures B and C. $\alpha$ and $\delta$ refers to the peaks identified in the luminosity weighted map.}
\label{tab:BC}
\begin{center}
\begin{tabular}{l c c c c c c c c c c}
\toprule

&	 	&   	&	& &  & \multicolumn{2}{c|}{$M_{200}-\sigma$} &\multicolumn{2}{c|}{Caustic}\\ \cmidrule{7-10}
        	
Subcluster	&	$\alpha$	&   $\delta$	&	$N_{\rm gals}$ & $\bar{z}$	&	$\sigma_v/(1+\bar{z})$	&	$M_{200}$	&	$R_{200}$	&		$M_{200}$	&	$R_{200}$	\\[5pt]
        	&	(J2000)  	&   (J2000)  	&	        	&		        &	(km s$^{-1}$)	&	($10^{14}$ M$_\odot$)	&	(kpc)	        & ($10^{13}$ M$_\odot$)	    & (kpc)	\\

\midrule

B 	 & 	 239.46532  	 & 	  16.31289  	 & 	  30  	 & 	  $  0.0369\pm   0.0012$  	 & 	  $356_{- 21}^{+ 21}$  	 & 	  $0.42_{-0.08}^{+0.09}$  	 & 	   $ 708_{-  51}^{+  47}$    	 & 	     $0.55_{-0.13}^{+0.15}$  	 & 	   $ 807_{-  65}^{+  70}$ \\[5pt]
C 	 & 	 239.59758  	 & 	  18.06715  	 & 	  41  	 & 	  $  0.0464\pm   0.0013$  	 & 	  $374_{- 24}^{+ 23}$  	 & 	  $0.49_{-0.11}^{+0.10}$  	 & 	   $ 745_{-  58}^{+  47}$    	 & 	    $0.82_{-0.27}^{+0.14}$  	 & 	   $ 925_{- 117}^{+  50}$ \\[5pt]

\bottomrule

\end{tabular}
\end{center}
\end{table*}

Within the error bars both methods present comparable estimates. The total mass of each cluster, considering the sum of its respective subclusters is $2.88_{-0.27}^{+0.31} \times 10^{14}$ M$_\odot$ for A2151, $13.5_{-1.7}^{+2.1} \times 10^{14}$ M$_\odot$ for A2147 and $0.72_{-0.10}^{+0.13} \times 10^{14}$ for M$_\odot$ A2152. These quantities were estimated by \cite{Barmby98} taking each cluster as a single structure. The virial masses were $7.0\pm0.9 \times 10^{14}$  M$_\odot$, $13.0\pm2.0 \times 10^{14}$  M$_\odot$, and $7.2\pm1.7 \times 10^{14}$  M$_\odot$ respectively for A2151, A2147 and A2152. Except for A2147, where there is a good agreement, the other two estimates are higher than ours.

The internal energy of interacting clusters changes during the lifetime of a merger, leading to a temporary modification in their PPS. This effect is commonly seen as a boost in the cluster velocity dispersion in periods close to the pericentric passage \cite[e.g.][]{pinkney,Takizawa10,Monteiro-Oliveira20}, meaning that the masses obtained have to be considered with wariness. In fact, the comparison of the dynamical-based masses with the lensing-based ones is often called an indicator of the cluster dynamical status \citep[e.g.][]{Soja18,Monteiro-Oliveira21}.

\section{Kinematic analysis}
\label{sec:two.body}

\begin{figure}
\begin{center}
\includegraphics[width=1.0\columnwidth, angle=0]{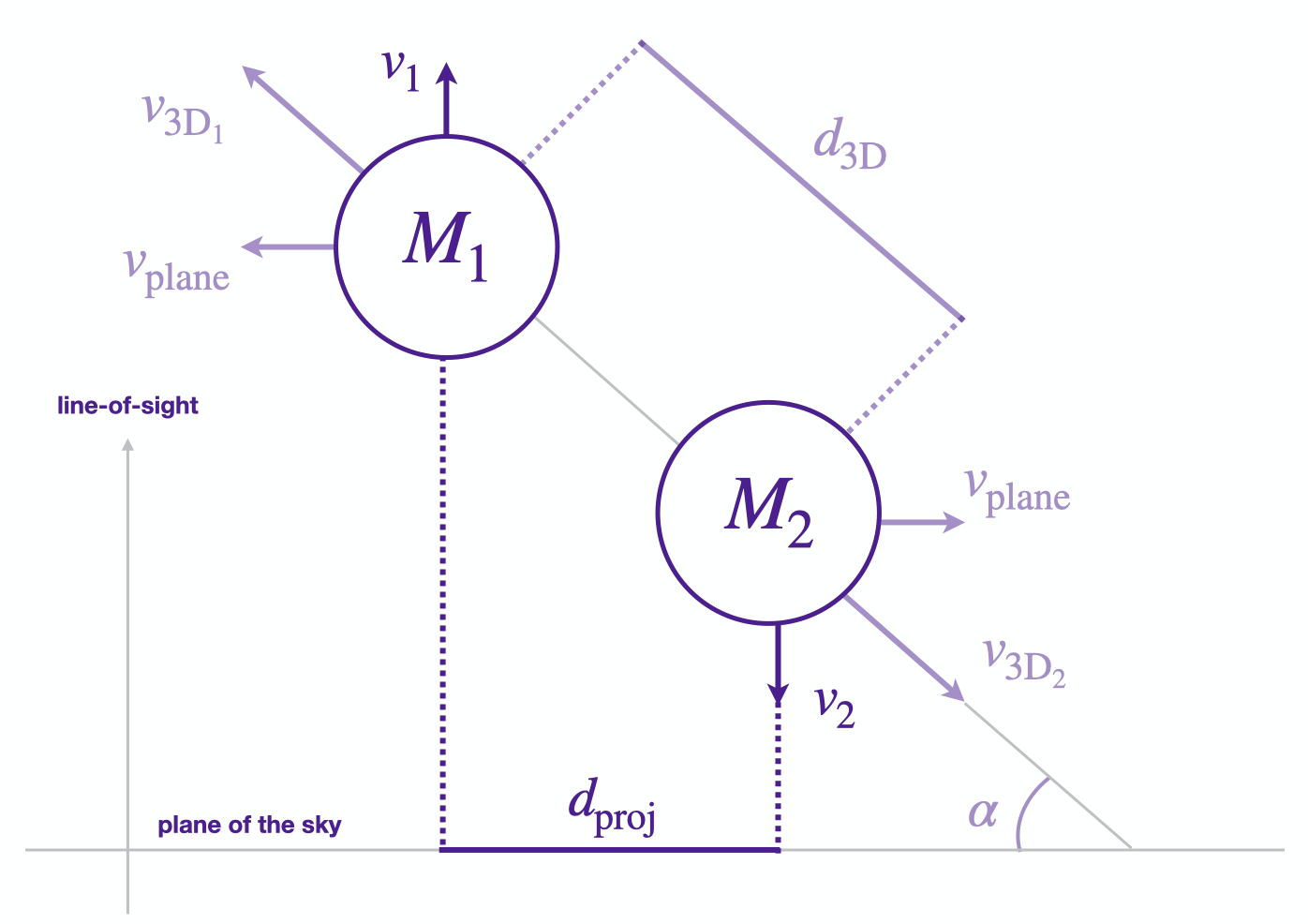}     
\caption{Typical configuration of a two-body system. The input quantities of MCMAC-post are highlighted: the halo masses ($M_{1}$ and $M_{2}$) and their mean redshift ($\bar{z}$) as well as and their projected separation ($d_{\rm proj}$). The corresponding 3-dimensional quantities are not directly accessible since we do not know the angle $\alpha$ between the merger axis and the plane of the sky.}
\label{fig:dawson.scheme}
\end{center}
\end{figure}

The Monte Carlo Merging Analysis Code (hereafter MCMAC-post) was introduced by \cite{dawson} and consists of an analytical description of the merger between two galaxy clusters.  It determines the dynamical solutions for bounded halos whose mass density follows NFW profiles \citep{nfw96,nfw97} truncated at $R_{200}$. The concentration parameter is fixed by the $M_{200}-c_{200}$ scaling relation given by \cite{duffy08}. For the sake of simplification, the collision is supposed to occur with no impact parameter and no angular momentum. However, a comparison with numerical simulations showed that all previous considerations do not play any significant effect on the final results \citep{dawson}. A typical merging configuration is illustrated in Fig.~\ref{fig:dawson.scheme}.

The code requires as input the halo masses, the projected distance between them\footnote{Computed as the separations of their respective luminosity-weighted peaks (Fig.~\ref{fig:peaks}).} and their corresponding mean redshift. From these parameters, the MCMAC-post make their probability density functions to generate plausible states of the two body's movement. A code extension presented in \cite{andrade-santos15} (hereafter MCMAC-pre) incorporate also unbound solutions for pre-merger systems, i.e., the left term of the motion equation:
\begin{equation}
V_r^2R_p \leq 2GM\sin^2 \alpha \cos \alpha,
\label{eq:bound}
\end{equation}
is now allowed to be greater than the right side, being $V_r$ the radial velocity between the clusters, $R_P$ their corresponding projected distance and $M$ the system total mass \citep{beers82}. However, the MCMAC-pre additionally requires that $V_r$ be less than the l.o.s. Hubble flow velocity. Then, the code defines the probability of a system to be (un)bound as the ratio between the sum of the respective states assumed by Equation~\ref{eq:bound} and the total number of realizations.

However, it is known that the uncertainties of the final results are relatively high because the exact geometry of the merger is not known a priori (angle $\alpha$ in Fig.~\ref{fig:dawson.scheme}). A way to overcome this issue is to compute $\alpha$ from the velocity components:

\begin{equation}
\alpha=\arctan \left ( \frac{\delta v}{v_{\rm plane}}\right).
\label{eq:theta}
\end{equation}
being $\delta v$ the difference in velocity between the two clusters along the l.o.s., which comes from the data. Therefore, constraining $v_{\rm plane}$, the relative velocity along the plane of the sky, we are directly restringing $\alpha$. Based on previous knowledge of cluster merger kinematics though hydrodynamical simulations \citep{springel07,machado13,Machado+2015,Doubrawa20,Moura20} we adopted a uninformative prior  $v_{\rm plane}$ < 1500 km s$^{-1}$.

Our first goal is to determine what pairs are dynamically bound to each other, two by two,  by running the MCMAC-pre. Then, we have run the MCMAC-post to unveil the kinematics of each merger. In both cases, we have considered the mass from $M_{200}-\sigma$ scaling relation as the fiducial. We will discuss the merger scenarios following. We reinforce that all forthcoming results remaining comparable when we have considered the caustic mass as input in the MCMAC.

\subsection{A2151}
\label{sec:two.body.A2151}

\begin{figure*}
\begin{center}
\includegraphics[width=1.0\textwidth, angle=0]{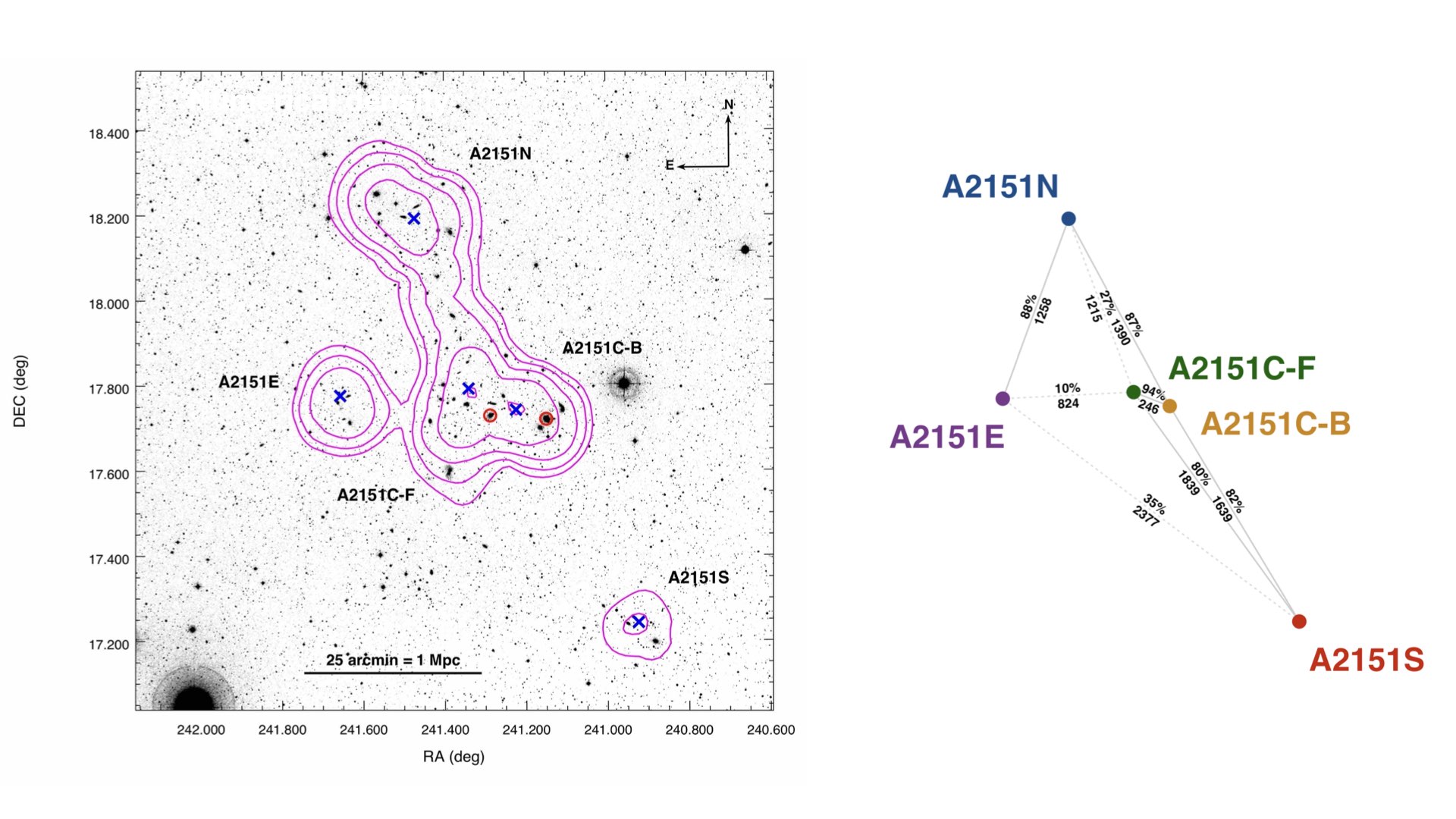}     
\caption{The galaxy cluster A2151 ($\bar{z}=0.0361\pm0.0024$, a.k.a. Hercules cluster) and its main substructures. {\it Left: } Luminosity weighted map of the spectroscopic members (magenta contours) overlaid with optical DSS2-red ESO Online Digitized Sky Survey image. We found five significative galaxy clumps in this region (blue crosses), corresponding to the A2151's subclusters. Only two of them (A2151C-B and A2151C-F) have a measurable gas counterparts, whose X-ray emission peaks are marked with red circles. {\it Right: } A2151 skeleton. The  respective distances among the clusters are showed in units of kpc. It is also presented the probability of each pair to be bound (filled lines) as was obtained by the MCMAC-pre code.}
\label{fig:2151.ske}
\end{center}
\end{figure*}

A summary of the A2151 structure is present in Fig.~\ref{fig:2151.ske}.  The probability of each of them be bound is shown in Fig.~\ref{fig:2151.ske}, right. Each of the four known components (N, E, C-B and C-F) are linked with at least one companion. Our analysis has shown that the southern clump (``D'' in Fig.~\ref{fig:maps}), is gravitationally bound with A2151C-F and A2151C-B, confirming, therefore, that A2151S is, in fact, part of the cluster A2151 as a whole.

With exception of A2151E, which is linked only with A2151N, the other subsclusters form, at least, two bounded pair each. A2151C-B is the most connected, being linked with A2151C-F, A2151S and A2151N. Noteworthy, is our find that A2151E and A2151C-F are not bounded. This conclusion excludes the previous conjecture of \cite{Bird95} in which both consists of a post-merger system.

Due to the model's inability to discern between the pre- and post-merging states, we will describe both scenarios for the bounded pairs. The only relatively unquestionable fact is the pre-merger state between the central substructures in the sense that the cool core of A2151C-B is still preserved \citep{Lagana19,Tiwari21}. The encounter will be happening in $0.38^{+0.12}_ {-0.20}$ Gyr with a velocity of $1147_{-139}^{+169}$ km s$^{-1}$ at a moderated distance of the plane of the sky (${46}_{-19}^{+22}$ degrees). At the apoapsis, both subclusters will be apart by  $1.47_{-1.2}^{+1.1}$ Mpc.

Other possible encounters would spend more than $1.6$ Gyr to take place (for example, A2151N and A2151E), which is at least  four times larger than the time of the collision between the central structures A2151C-B and A2151C-F. Given this scenario, all other models involving one of the central subclusters consists of a toy model, since their dynamical state is going to change faster in comparison to the outskirts subclusters.

\begin{landscape}
\begin{table}
\caption[]{Dynamical description of the bound and pre-merger pairs according to MCMAC-pre output parameters. $M$ is the cluster mass, $z$ the mean redshift, $d_{\rm proj}$ the observed projected density, $v_{\rm rad, obs}$ the observed relative radial velocity, $\alpha$ the merger angle, $v_{\rm 3D,obs}$ the 3D observed velocity, $d_{\rm 3D,obs}$ the current 3D distance, $d_{\rm 3D,max}$ the distance at apoapsis, and $TTC$ is the time until the collision.}
\label{tab:A2151.pre}
\begin{center}
\begin{tabular}{l c | c c | c c | c c | c c | c c}
\toprule
\multicolumn{12}{c}{Abell 2151}\\
\toprule

\multicolumn{2}{l|}{Pre-merger pairs} & \multicolumn{2}{c|}{C-B(1) -- C-F(2)} & \multicolumn{2}{c|}{S(1) -- C-F(2)} & \multicolumn{2}{c|}{S(1) -- C-B(2)} & \multicolumn{2}{c}{N(1) -- E(2)}  & \multicolumn{2}{c}{N(1) -- C-B(2)}\\  \cmidrule{3-12}

\multicolumn{2}{l|}{} &  Median	&	68 per cent c.l &  Median	&	68 per cent c.l&  Median	&	68 per cent c.l&  Median	&	68 per cent c.l  &  Median	&	68 per cent c.l \\

\midrule

$M_{\rm (1)}$	  &  	$10^{14}$ M$_\odot$	  &  	0.45	 & 	0.36 -- 0.55	   &  	1.00	 & 	0.80 -- 1.17	    &  	0.99	 & 	0.81 -- 1.19	    &  	0.63	 & 	0.50 -- 0.74	    &  	0.63	 & 	0.52 -- 0.75	   \\ 
$M_{\rm (2)}$	  &  	$10^{14}$ M$_\odot$	  &  	0.20	 & 	0.17 -- 0.25	   &  	0.20	 & 	0.17 -- 0.24	    &  	0.44	 & 	0.34 -- 0.54	    &  	0.66	 & 	0.54 -- 0.77	    &  	0.44	 & 	0.34 -- 0.54	   \\ 
$z_{\rm (1)}$	  &  	--	  &  	0.0352	 & 	0.0343 -- 0.0360	   &  	0.0333	 & 	0.0323 -- 0.0342	    &  	0.0353	 & 	0.0341 -- 0.0363	    &  	0.0380	 & 	0.0369 -- 0.0390	    &  	0.0370	 & 	0.0359 -- 0.0379	   \\ 
$z_{\rm (2)}$	  &  	--	  &  	0.0334	 & 	0.0325 -- 0.0341	   &  	0.0330	 & 	0.0321 -- 0.0337	    &  	0.0357	 & 	0.0347 -- 0.0367	    &  	0.0384	 & 	0.0374 -- 0.0395	    &  	0.0367	 & 	0.0357 -- 0.0376	   \\ 
$d_{\rm proj}$	  &  	Mpc	  &  	0.25	 & 	0.20 -- 0.29	   &  	1.84	 & 	1.80 -- 1.89	    &  	1.64	 & 	1.59 -- 1.69	    &  	1.26	 & 	1.21 -- 1.31	    &  	1.22	 & 	1.18 -- 1.27	   \\ 
$v_{\rm rad, obs}$	  &  	km s$^{-1}$	  &  	 547	 & 	 348 --  801	   &  	 182	 & 	   1 --  259	    &  	 217	 & 	   1 --  305	    &  	 235	 & 	   2 --  327	    &  	 203	 & 	   0 --  287	   \\ 
$\alpha$	  &  	degrees	  &  	  46	 & 	  27 --   68	   &  	  29	 & 	   0 --   41	    &  	  30	 & 	   0 --   43	    &  	  30	 & 	   0 --   43	    &  	  28	 & 	   0 --   41	   \\ 
$v_{\rm 3D,obs}$	  &  	km s$^{-1}$	  &  	 828	 & 	 650 -- 1121	   &  	 449	 & 	 291 --  676	    &  	 522	 & 	 329 --  765	    &  	 549	 & 	 361 --  833	    &  	 511	 & 	 300 --  750	   \\ 
$d_{\rm 3D,obs}$	  &  	Mpc	  &  	0.36	 & 	0.21 -- 0.48	   &  	2.10	 & 	1.75 -- 2.47	    &  	1.89	 & 	1.55 -- 2.25	    &  	1.46	 & 	1.19 -- 1.75	    &  	1.38	 & 	1.15 -- 1.65	   \\ 
$v_{\rm 3D,col}$	  &  	km s$^{-1}$	  &  	1147	 & 	1008 -- 1316	   &  	1564	 & 	1468 -- 1666	    &  	1603	 & 	1510 -- 1709	    &  	1508	 & 	1406 -- 1609	    &  	1423	 & 	1333 -- 1526	   \\ 
$d_{\rm 3D,max}$	  &  	Mpc	  &  	1.47	 & 	0.27 -- 2.57	   &  	4.99	 & 	1.84 -- 8.14	    &  	4.56	 & 	1.62 -- 7.60	    &  	3.38	 & 	1.25 -- 5.51	    &  	3.17	 & 	1.22 -- 5.14	   \\ 
$TTC$	  &  	Gyr	  &  	0.38	 & 	0.18 -- 0.50	   &  	2.77	 & 	1.66 -- 3.54	    &  	2.20	 & 	1.34 -- 2.87	    &  	1.63	 & 	0.98 -- 2.09	    &  	1.66	 & 	0.99 -- 2.11	   \\

\toprule
\toprule

\multicolumn{12}{c}{Abell 2152 + A2147 + A}\\
\toprule

\multicolumn{2}{l}{Pre-merger pairs} & \multicolumn{2}{c}{A2147S(1) -- A2147N(2)} & \multicolumn{2}{c}{A2152S(1) -- A2152N(2)} & \multicolumn{2}{c}{A(1) -- A2147S(2)} & 
\multicolumn{2}{c}{A(1) -- A2152S(2)}  & 
\multicolumn{2}{c}{}\\  \cmidrule{3-12}

\multicolumn{2}{l|}{} &  Median	&	68 per cent c.l &  Median	&	68 per cent c.l&  Median	&	68 per cent c.l&  Median	&	68 per cent c.l  &  	&	 \\

\midrule

$M_{\rm (1)}$	  &  	$10^{14}$ M$_\odot$	  &  	3.45	 & 	2.73 -- 4.17	   &  	0.24	 & 	0.19 -- 0.29	    &  	3.15	 & 	2.58 -- 3.74	    &  	3.22	 & 	2.64 -- 3.80	    &  &    \\ 
$M_{\rm (2)}$	  &  	$10^{14}$ M$_\odot$	  &  	10.19	 & 	8.11 -- 12.08	   &  	0.49	 & 	0.38 -- 0.58	    &  	3.52	 & 	2.79 -- 4.19	    &  	0.24	 & 	0.19 -- 0.29	    &  &    \\ 
$z_{\rm (1)}$	  &  	--	  &  	0.0358	 & 	0.0335 -- 0.0379	   &  	0.0444	 & 	0.0436 -- 0.0452	    &  	0.0383	 & 	0.0363 -- 0.0399	    &  	0.0425	 & 	0.0411 -- 0.0437	    &   &    \\ 
$z_{\rm (2)}$	  &  	--	  &  	0.0366	 & 	0.0341 -- 0.0392	   &  	0.0447	 & 	0.0439 -- 0.0457	    &  	0.0373	 & 	0.0353 -- 0.0389	    &  	0.0436	 & 	0.0427 -- 0.0446	    &   &    \\ 
$d_{\rm proj}$	  &  	Mpc	  &  	1.19	 & 	1.14 -- 1.23	   &  	0.93	 & 	0.88 -- 0.98	    &  	2.53	 & 	2.48 -- 2.57	    &  	1.74	 & 	1.69 -- 1.79	    &  &    \\ 
$v_{\rm rad, obs}$	  &  	km s$^{-1}$	  &  	 542	 & 	   1 --  821	   &  	 193	 & 	   0 --  273	    &  	 424	 & 	  12 --  581	    &  	 383	 & 	 114 --  612	    &   &   \\ 
$\alpha$	  &  	degrees	  &  	  45	 & 	  21 --   73	   &  	  28	 & 	   0 --   41	    &  	  32	 & 	   0 --   45	    &  	  34	 & 	   3 --   48	    &   &   \\ 
$v_{\rm 3D,obs}$	  &  	km s$^{-1}$	  &  	 947	 & 	   3 -- 1385	   &  	 480	 & 	 291 --  717	    &  	 902	 & 	 578 -- 1295	    &  	 789	 & 	 513 -- 1150	    &   &   \\ 
$d_{\rm 3D,obs}$	  &  	Mpc	  &  	1.69	 & 	1.12 -- 2.32	   &  	1.06	 & 	0.85 -- 1.26	    &  	3.00	 & 	2.45 -- 3.58	    &  	2.10	 & 	1.65 -- 2.52	    &  &    \\ 
$v_{\rm 3D,col}$	  &  	km s$^{-1}$	  &  	2830	 & 	2376 -- 3351	   &  	1250	 & 	1153 -- 1348	    &  	2601	 & 	2443 -- 2757	    &  	2251	 & 	2090 -- 2428	    &   &   \\ 
$d_{\rm 3D,max}$	  &  	Mpc	  &  	2.75	 & 	1.15 -- 4.31	   &  	2.45	 & 	0.91 -- 3.90	    &  	7.43	 & 	2.55 -- 12.42	    &  	5.46	 & 	1.71 -- 9.14	    &  &    \\ 
$TTC$	  &  	Gyr	  &  	0.99	 & 	0.38 -- 1.35	   &  	1.38	 & 	0.82 -- 1.76	    &  	2.03	 & 	1.22 -- 2.59	    &  	1.65	 & 	0.95 -- 2.14	    &  &    \\ 

\bottomrule

\end{tabular}
\end{center}
\end{table}

\end{landscape}

As justified before, we have excluded all pairs containing A2151C-B before running MCMAC-post. After that, only two pairs have remained: N--E and S--C-F. It is important to remark that the post-merger scenario described by the code is degenerated because it cannot point which outgoing or incoming state is the most likely. Usually this degeneracy can be broken with multi-wavelength observations \citep[e.g.][]{ng15,Monteiro-Oliveira17b,Kim21}. However, our results point into a very unlikely incoming scenario for both pairs, because the time since the last collision ($TSC_1$) is very large compared to the Hubble time. Therefore, this sketch was ruled out.

The outgoing post-merger scenario for A2151N--A2151E predicts that they had a pericentric passage $1.6_{-0.6}^{+0.4}$ Gyr ago with $1505_{-92}^{+108}$ km s$^{-1}$  with a merger axis-aligned $28_{-28}^{+13}$ degrees in respect to the plane of the sky. The system would have already covered $\sim$43 per cent of the path to the apoapsis where they can be found $3.3_{-2.1}^{+2.0}$ Mpc apart. For the system A2151S--A2151C-F, the encounter happened $2.8_{-1.1}^{+0.7}$ Gyr ago with $1563_{-94}^{+101}$ km s$^{-1}$ which will lead to a maximum separation of $4.8_{-2.9}^{+2.8}$ Mpc, whose $\sim$44 per cent have already been covered.

\begin{table*}
\caption[]{Dynamical description of the post-merger pairs. Most of the parameters are the same as those in Table~\ref{tab:A2151.pre}, but now including $TSC_0$,  the time since the pericentric passage for an outcoming system, $TSC_1$ is the time since the pericentric passage for an incoming system,  $T$  the period between two successive collisions, and $P$ the probability of observing the system.}
\label{tab:A2151.post}
\begin{center}
\begin{tabular}{l c | c c | c c }
\toprule
\multicolumn{6}{c}{Abell 2151}\\
\toprule

\multicolumn{2}{l|}{Post-merger pairs}  & \multicolumn{2}{c|}{N(1) -- E(2)} & \multicolumn{2}{c|}{S(1) -- C-F(2)} \\  \cmidrule{3-6}

\multicolumn{2}{l|}{} &  Median	&	68 per cent c.l &  Median	&	68 per cent c.l\\

\midrule

$M_{\rm (1)}$	  &  	$10^{14}$ M$_\odot$	  &  	0.63	 & 	0.51 -- 0.74	   &  	1.00	 & 	0.81 -- 1.18	    \\ 
$M_{\rm (2)}$	  &  	$10^{14}$ M$_\odot$	  &  	0.66	 & 	0.54 -- 0.77	   &  	0.20	 & 	0.16 -- 0.24	    \\ 
$z_{\rm (1)}$	  &  	--	  &  	0.0380	 & 	0.0369 -- 0.0390	   &  	0.0333	 & 	0.0323 -- 0.0343	   \\ 
$z_{\rm (2)}$	  &  	--	  &  	0.0384	 & 	0.0375 -- 0.0395	   &  	0.0330	 & 	0.0322 -- 0.0338	   \\ 
$d_{\rm proj}$	  &  	Mpc	  &  	1.26	 & 	1.21 -- 1.31	   &  	1.84	 & 	1.79 -- 1.88	    \\ 
$v_{\rm rad, obs}$	  &  	km s$^{-1}$	  &  	 229	 & 	   0 --  323	   &  	 188	 & 	   2 --  266	  \\ 
$\alpha$	  &  	degrees	  &  	  28	 & 	   0 --   41	   &  	  28	 & 	   0 --   40	  \\ 
$v_{\rm 3D,obs}$	  &  	km s$^{-1}$	  &  	 558	 & 	 345 --  807	   &  	 456	 & 	 288 --  649	  \\ 
$d_{\rm 3D,obs}$	  &  	Mpc	  &  	1.43	 & 	1.17 -- 1.68	   &  	2.09	 & 	1.77 -- 2.45	    \\ 
$v_{\rm 3D,col}$	  &  	km s$^{-1}$	  &  	1505	 & 	1413 -- 1614	   &  	1563	 & 	1469 -- 1664	  \\ 
$d_{\rm 3D,max}$	  &  	Mpc	  &  	3.29	 & 	1.23 -- 5.27	   &  	4.77	 & 	1.87 -- 7.56	    \\ 
$TSC_0$	  &  	Gyr	  &  	1.60	 & 	0.97 -- 2.01	   &  	2.75	 & 	1.65 -- 3.45	    \\ 
$TSC_1$	  &  	Gyr	  &  	15.57	 & 	2.27 -- 32.91	   &  	28.08	 & 	3.94 -- 58.14	    \\ 
$T$	  &  	Gyr	  &  	17.71	 & 	4.20 -- 35.56	   &  	31.34	 & 	7.66 -- 62.48	    \\ 
$P$	  &  	per cent	  &  	  21	 & 	   0 --   35	   &  	  19	 & 	   0 --   33	  \\

\toprule

\multicolumn{2}{l|}{Post-merger pairs}  & \multicolumn{2}{c|}{A2147S(1) -- A2147N(2)} & \multicolumn{2}{c|}{A2152S(1) -- A2152N(2)} \\  \cmidrule{3-6}

\multicolumn{2}{l|}{} &  Median	&	68 per cent c.l &  Median	&	68 per cent c.l\\

\midrule

$M_{\rm (1)}$	  &  	$10^{14}$ M$_\odot$	  &  	3.46	 & 	2.75 -- 4.16	   &  	0.24	 & 	0.19 -- 0.29	    \\ 
$M_{\rm (2)}$	  &  	$10^{14}$ M$_\odot$	  &  	10.20	 & 	8.38 -- 12.35	   &  	0.49	 & 	0.39 -- 0.59	    \\ 
$z_{\rm (1)}$	  &  	--	  &  	0.0356	 & 	0.0334 -- 0.0377	   &  	0.0444	 & 	0.0436 -- 0.0453	   \\ 
$z_{\rm (2)}$	  &  	--	  &  	0.0368	 & 	0.0343 -- 0.0398	   &  	0.0447	 & 	0.0438 -- 0.0456	   \\ 
$d_{\rm proj}$	  &  	Mpc	  &  	1.19	 & 	1.14 -- 1.24	   &  	0.93	 & 	0.88 -- 0.98	    \\ 
$v_{\rm rad, obs}$	  &  	km s$^{-1}$	  &  	 698	 & 	   3 --  984	   &  	 194	 & 	   0 --  276	  \\ 
$\alpha$	  &  	degrees	  &  	  43	 & 	  19 --   68	   &  	  28	 & 	   0 --   40	  \\ 
$v_{\rm 3D,obs}$	  &  	km s$^{-1}$	  &  	1204	 & 	 666 -- 1668	   &  	 490	 & 	 304 --  722	  \\ 
$d_{\rm 3D,obs}$	  &  	Mpc	  &  	1.63	 & 	1.11 -- 2.05	   &  	1.06	 & 	0.86 -- 1.25	    \\ 
$v_{\rm 3D,col}$	  &  	km s$^{-1}$	  &  	2710	 & 	2200 -- 3083	   &  	1249	 & 	1151 -- 1344	  \\ 
$d_{\rm 3D,max}$	  &  	Mpc	  &  	2.32	 & 	1.20 -- 3.43	   &  	2.39	 & 	0.91 -- 3.73	    \\ 
$TSC_0$	  &  	Gyr	  &  	0.83	 & 	0.56 -- 1.01	   &  	1.36	 & 	0.81 -- 1.72	    \\ 
$TSC_1$	  &  	Gyr	  &  	2.95	 & 	1.05 -- 5.06	   &  	12.73	 & 	1.89 -- 25.70	    \\ 
$T$	  &  	Gyr	  &  	3.80	 & 	2.00 -- 6.30	   &  	14.56	 & 	3.51 -- 27.77	    \\ 
$P$	  &  	per cent	  &  	  43	 & 	  18 --   68	   &  	  21	 & 	   0 --   36	  \\ 

\bottomrule

\end{tabular}
\end{center}
\end{table*}

\subsection{A2147 + A2152 + A}
\label{sec:two.body.A2152}

\begin{figure*}
\begin{center}
\includegraphics[width=1.0\textwidth, angle=0]{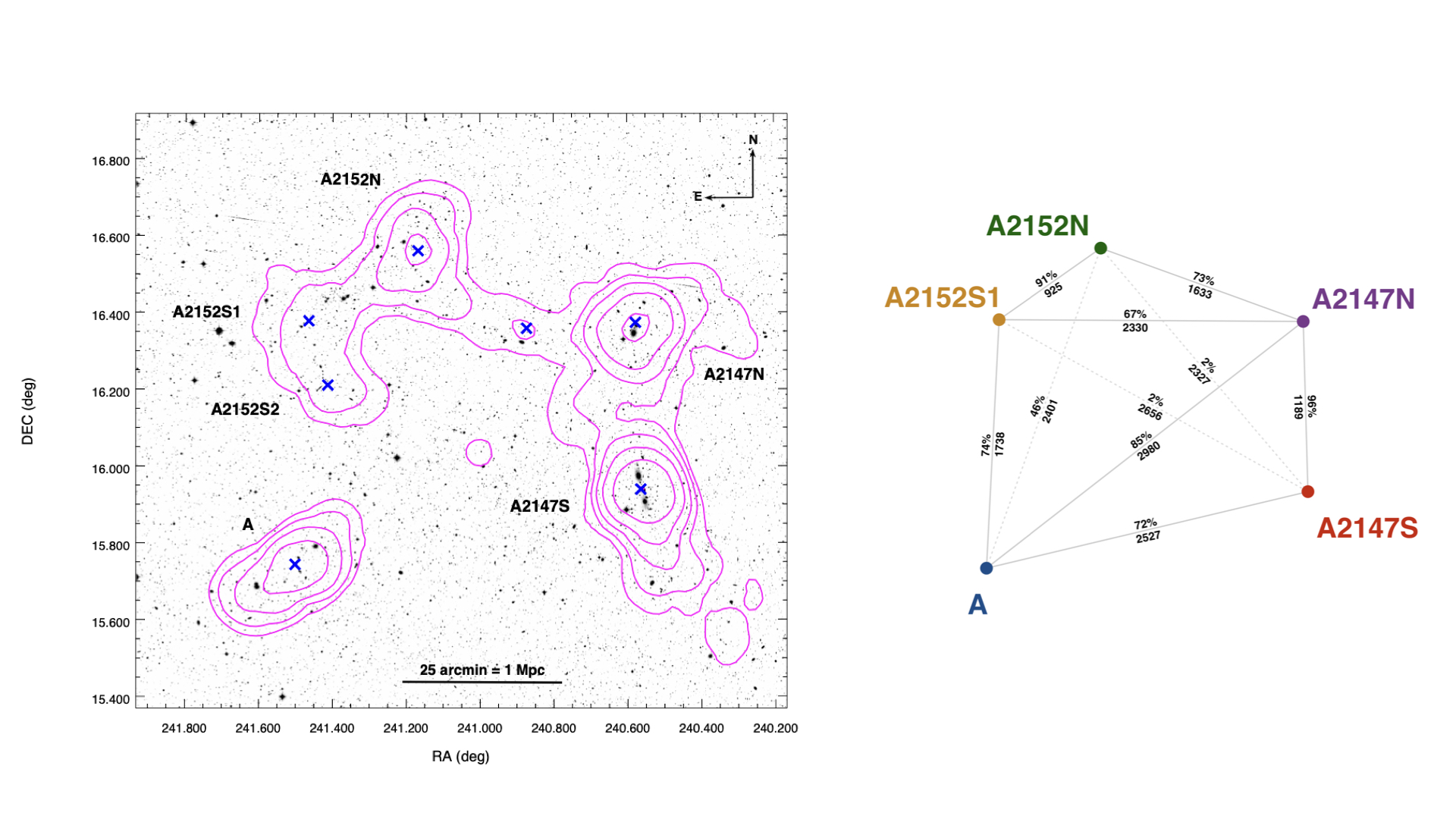}     
\caption{Same as Fig.~\ref{fig:2151.ske}, but now showing the region of A2147 + A2152 + A.}
\label{fig:2147.ske}
\end{center}
\end{figure*}

A zoomed view and the dynamical scheme is shown in Fig.~\ref{fig:2147.ske}. The clusters A2147 and A2152 have their internal substructures strongly bounded, with a probability larger than 90 per cent. The subcluster A2147N is the only one bounded with all other companions,  which can be interpreted as a consequence of its dominant mass in the field. Regarding mass clump A, we confirm that it is part of the Hercules supercluster, being bounded with both clusters.

We will concentrate our discussion on the internal dynamic of each cluster and the interactions involving the clump A and their nearest neighbours (A2147S and A2152S1). In the pre-merger scenario, the subclusters of A2147 will have their pericentric passage in $0.99_{-0.68}^{+0.36}$ Gyr with a velocity of $2830_{-454}^{+521}$ km s$^{-1}$ at a moderate collision axis ($45_{-24}^{+28}$ degrees). For A2152, the values are respectively,  $1.38_{-0.56}^{+0.38}$ Gyr, $1250_{-97}^{+98}$ km s$^{-1}$, and $28_{-28}^{+13}$ degrees. The possible collision of A with A2147S would happen in $2.03_{-0.81}^{+0.56}$ Gyr, with $2601_{-158}^{+156}$ km s$^{-1}$, and  $32_{-32}^{+13}$ degrees, whereas the possible collision with A2152S1 is characterized by $1.65_{-0.70}^{+0.49}$ Gyr,  $2251_{-161}^{+177}$ km s$^{-1}$, and  $34_{-31}^{+14}$ degrees.


The elongated shape of X-ray emission in A2147 \citep[e.g.][]{Sanderson06,Vikhlinin09} is a hint that the system already experienced a collision. In this case, if the system has been caught outgoing, the collision happened $0.83_{-0.27}^{+0.18}$ Gyr ago, with $2710_{-510}^{+373}$ km s$^{-1}$, and  $43_{-24}^{+25}$ degrees having the clusters travelled a distance equivalent to 70 per cent of the path for the maximum separation at $2.32_{-1.12}^{+1.11}$ Mpc. In the incoming scenario, the encounter would have occurred $2.95_{-1.90}^{+2.11}$ Gyr ago. 

The dynamical description of A2152 is a little bit less complicated because the incoming scenario can be disregarded since its related time is comparable to the Hubble one. Then, in case of the subclusters has already collided, this event happened  $1.36_{-0.55}^{+0.36}$ Gyr ago with an encounter velocity of $1249_{-98}^{+95}$ km s$^{-1}$ along a merger axis located $28_{-28}^{+12}$ degrees from the plane of the sky. The subclusters have toured only 44 per cent of their path to reach the appoasis at $2.39_{-1.48}^{+1.34}$ Mpc.

\section{Discussion}
\label{sec:discussion}

\subsection{Merger impact on dynamical mass estimation}
\label{sec:merger.impact}

The merger of galaxy clusters involves energies up to $10^{64}$ ergs \citep{sarazin04} being part of this amount converted into internal movement of the galaxies. These energy transfer change the PPS leading to a temporary boost in the velocity dispersion near the pericentric passage  \citep{pinkney}. Hence, the question we ask is: how do this ephemeral disturbance bias the dynamical based mass estimations? Nevertheless, observations of a particular system only afford the description of a specific snapshot of the whole merger process.  A proper answer to the question is not a simple task and the use of realistic computational simulations is a first step in addressing the issue as they provide a follow-up over the time of the interaction between the clusters \citep[e.g.][]{Roettiger96}.

\begin{figure}
\begin{center}
\includegraphics[width=1.0\columnwidth, angle=0]{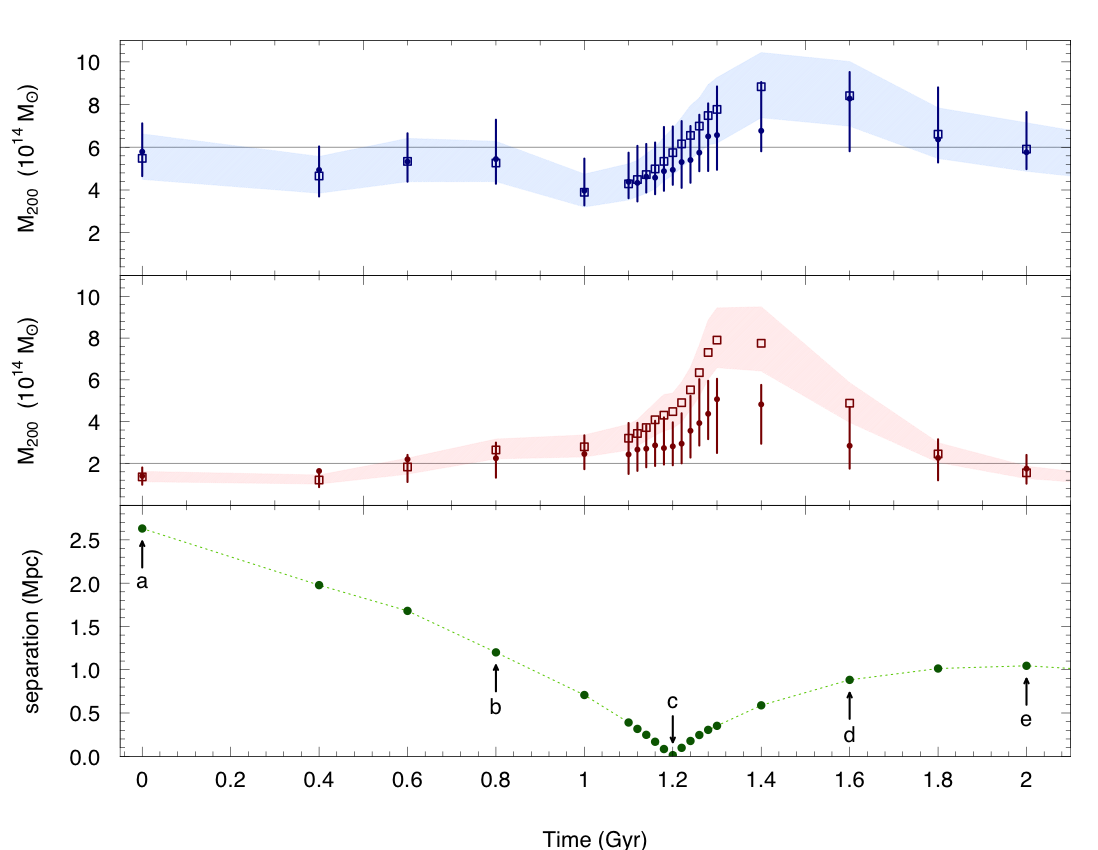} 
\caption{Evolution of the mass estimators across the timeline of a 1:3 mass ratio merger, with null impact parameter and a collision axis parallel to the plane of the sky. Main cluster ($6 \times 10^{14}$ M$_\odot$) and the subcluster ($2 \times 10^{14}$ M$_\odot$) are respectively in the upper (blue) and middle (red) panels. The $M_{200}-\sigma$ value is represented by squares and the respective errors by a shadow area whereas the caustic masses are marked by dots with error bars. In the bottom, we can see the clusters separation. We have defined five particular moments: ($a$) initial, ($c$) periapsis, ($e$) apoapsis and two intermediate points ($b$ and $d$), in which we compared the MCMAC predictions with the simulation results. A clear boost in the dynamical mass regardless the method  can be seen between the instants $c$ and $e$.}
\label{fig:zuhone}
\end{center}
\end{figure}

To address this inquiry, we have resorted to the ``Galaxy Cluster Merger Catalog'', a suite of N-body and hydrodynamical simulations made publicly available by \cite{Zuhone16}. The rich data set provides multi-wavelength high-resolution simulations of the merger between two clusters over a timeline of 10 Gyr. To accomplish our goal and access the merger kinematic, we have considered the galaxy catalogue represented by dark matter particles of the simulation (projected position and redshift). Among the several configurations, we choose the merger having a mass ratio of 1:3 (2:6 $\times 10^{14}$ M$_\odot$), with a null impact parameter ($b =0$ kpc), and whose merger axis is parallel to the plane of the sky ($\alpha=0$ degrees). This set up covers most of the mergers we have characterized in Secs.~\ref{sec:masses} and \ref{sec:two.body}. For each snapshot, we have computed the masses in the same fashion as done in Sec.~\ref{sec:masses} using all the respective galaxies belonging to each cluster (i.e. the masses were measured from the same galaxies along the process). To check the reliability of this assumption, we tracked the galaxies inside $R_{200}$, and we have found that the numerical variation is  less $\sim6\%$ for the major cluster and $\sim1\%$ in the minor, meaning an equivalent impact on the masses \citep[e.g.][]{Chiu20}. The resulting timeline is presented in Fig.~\ref{fig:zuhone}.

As expected, the mass estimated from the velocity dispersion (Eq.~\ref{eq:M.sigma}; squares and shadow areas in the plot) shows a boost in a period between the periapsis ($c$) and the apoapsis ($e$). However, another unexpected fact is observed: the caustic-based estimation is also biased through the merger age. This find is not in line with what is largely accepted in literature, that the caustic method is not dependent on the cluster dynamical state \citep[e.g.][]{Diaferio05,Geller13}. According to our results, the caustic mass behaviour is very similar to those from $\sigma$-based mass, with subtle differences, though. In the less massive cluster, there is a suggestion that the use of the caustic method would provide a less biased measure even during the most acute merger phase (1--2 Gyr). However, even in this case, the estimated masses are $\sim2.5\times$ higher than the real one but still lower than those from $M_{200}-\sigma$,  which are biased $\sim4\times$.

It is beyond the scope of this work to provide an ultimate explanation for the observed behaviour of the caustic masses. Consequently, our take-home message is that both dynamical mass estimators are subjected to a boost near to the pericentric passage and any kinematic analysis have to properly deal with this. However, we can speculate about the origin of caustic mass bias. Traditionally, to test its theoretical prediction of independence of the dynamical state, the caustic estimates are compared with the weak lensing based masses, this one unquestionably free of any equilibrium assumption. For example, considering a sample of three galaxy clusters,  \cite{Diaferio05} found a good agreement between both methods. With a larger sample (19), \cite{Geller13} reached the same conclusion. However, a scrutiny in their dataset reveals that the majority (10) are  clusters in some degree of interaction: A267 \citep[subsctrucutred;][]{Tucker20}, A1750 \citep[pre-merger;][]{Molnar13}, A1758 \citep[post-merger;][]{Machado+2015,Monteiro-Oliveira17a}, A1763 \citep[sloshing;][]{Douglass18}, A1835 \citep[out of hydrostatic equilibrium;][]{Ichikawa13},  A1914 \citep[out of hydrostatic equilibrium;][]{Mandal19}, A2034 \citep[post-merger;][]{Monteiro-Oliveira18,Moura20}, A2142 \citep[sloshing;][]{Rossetti13}, A2219 \citep[major merger;][]{Canning17}, A2631 \citep[late stage of merger;][]{Monteiro-Oliveira21}. Considering a multimodal structure as a single one will introduce a bias in the final mass leading, therefore to an unfruitful comparison.
Another concern, recently presented by \cite{Chadayammuri21}, is that the halo concentration is another quantity susceptible to changes near the pericentric passage. This means that weak lensing masses assuming a parameterized NFW model can present overestimated results because the model concentration assumes the values of relaxed clusters. Given this puzzling scenario, a more diligent comparison between caustic and weak lensing masses have to consider also the cluster dynamical state and/or the current merger phase.

Overall, it is important to state that our comparison with hydrodynamical simulations points that the bias in the caustic based mass occurs only during a short period of the cluster life. Furthermore, the caustic technique still provides us with a confident estimate of the dynamical mass, even in the case of the cluster is interacting and possibly far from the equilibrium state.

At last, we have compared our mass estimations with those available in the literature. The mass estimated by  \cite{Tiwari21} for A2151C-B is $1.3\pm0.8 \times 10^{14}$ M$_\odot$ and for A2151C-F is $0.42\pm0.30 \times 10^{14}$ M$_\odot$. Both are consistent with our results within their error bars. As stated in the introduction, no mass estimates were found for A2147 and A2152's subclusters, so we will compare them with the cluster total mass. \citep{Babyk14}, using {\it Chandra} data, have found $10.93_{-1.28}^{+1.27} \times 10^{14}$ M$_\odot$ for A2147  which is comparable with our estimate of the sum of the clusters constitutes. In case of A2152, \cite{Piffaretti11} found $0.81 \times 10^{14}$ M$_\odot$ also in a good agreement with our findings.

\subsection{Accuracy of the  kinematic description}
\label{sec:merger.description}

The modelling of interactions among $N>2$ bodies is an expensive task even from the computational point of view. This reflects in the small number of available studies involving more than two clusters  \citep[e.g.][]{Bruggen12,Ruggiero19,Doubrawa20}.
An adopted strategy to overcome such complexity is to start from a bimodal description and then to add a body acting as a ``perturber'' \citep[e.g.][]{Doubrawa20}. In the specific case of the present work, we can ask if describing a complex system as a composition of interactions two by two could give us at least, a general idea of the chronological order of the collisions, as well as determine which bodies are bound to the others. 

As the fiducial case, we have considered the triple merger in 1RXS J0603.3+4214 simulated by \cite{Bruggen12}. The initial conditions of the merger among the subclusters \#1, \#2 and \#3 are presented in their Table~1. We have used this information as input for the MCMAC-pre. The slight difference though is the impact parameter that originally ranges from 300--500 kpc but was assumed as null by the MCMAC-pre, as described in Sec.~\ref{sec:two.body}. The original simulation points that the collision between \#1 and \#2 happens $\sim1.3$ Gyr after simulation starts in $t = 0$ Gyr. Before that, however, a collision between \#2 and \#3 has happened at an unknown time.  The MCMAC-pre successfully recovers the bound state of the clusters. It is also efficient in describing the collision timeline, stating that  \#2 -- \#3 will collide first in 0.7 Gyr -- 1.4 Gyr and then  \#1 -- \#2 in  1.2 Gyr -- 2.3 Gyr, all within 68 per cent c.l. Of course, the simplicity of the approach does not allow us to describe the full history of collisions, but it gives a good overview of the initial events and their timeline.

Another fundamental question we can ask, is if our kinematic analysis is trustful given the merger impact on the mass determination. Or, in other words, does the MCMAC recover the correct physics of the merger even having as input possible biased masses? To address this question, we have chosen five snapshots in the ``Galaxy Cluster Merger Catalog'' subset described in Sec.~\ref{sec:merger.impact} (labelled $a$--$e$ in Fig.~\ref{fig:zuhone}). Then, we calculated the (sub)cluster masses (via scaling relation, as did in Sec.~\ref{sec:masses}) and used them as input for MCMAC. In the end, we have compared the outputs with the simulation predictions. The results can be seen in Table~\ref{tab:ZuHone.Dawson}.

\begin{table}
\caption[]{Comparison between the simulation predictions with the scenario recovered by MCMAC in five specific moments. We also considered the MCMAC results with and without a prior in the relative clusters velocity along the plane of the sky.}
\label{tab:ZuHone.Dawson}
\begin{center}
\begin{tabular}{l c | c | c c }
\toprule
Quantity & Unit & Sim. & MCMAC (w/ prior) & MCMAC (w/o prior) \\
 &  &  & 68 per cent c.l. & 68 per cent c.l. \\
\midrule
\multicolumn{5}{c}{$a$}\\
\midrule

$\alpha$	  &  deg	  &  0	  &  	 0 -- 37	   & 	12 -- 65	   \\ 
$d_{\rm 3D,max}$	  &   Mpc	  &  2.6	  &  	2.6 -- 8.8	   & 	2.6 -- 8.6	   \\ 
$TTC$	  &   Gyr	  &  1.2	  &  	1.2 -- 2.5	   & 	1.2 -- 5.0	   \\ 

\midrule
\multicolumn{5}{c}{$b$}\\
\midrule

$\alpha$	  &  deg	  &  0	  &  	 0 -- 39	   & 	 8 -- 62	   \\ 
$d_{\rm 3D,max}$	  &   Mpc	  &  2.6	  &  	1.2 -- 2.4	   & 	1.1 -- 3.2	   \\ 
$TTC$	  &   Gyr	  &  0.4	  &  	0.6 -- 1.1	   & 	0.5 -- 1.7	   \\ 

\midrule
\multicolumn{5}{c}{$c$}\\
\midrule

$\alpha$	  &  deg	  &  0	  &  	 0 -- 42	   & 	 3 -- 59	   \\ 
$d_{\rm 3D,max}$	  &   Mpc	  &  1.0	  &  	0.2 -- 0.7	   & 	0.0 -- 0.6	   \\ 
$TTC$	  &   Gyr	  &  0	  &  	0.0 -- 0.1	   & 	0.0 -- 0.3	   \\ 

\midrule
\multicolumn{5}{c}{$d$}\\
\midrule

$\alpha$	  &  deg	  &  0	  &  	 0 -- 35	   & 	 4 -- 55	   \\ 
$d_{\rm 3D,max}$	  &   Mpc	  &  1.0	  &  	0.9 -- 1.5	   & 	0.8 -- 2.0	   \\ 
$TSC_0$	  &   Gyr	  &  0.4	  &  	0.5 -- 0.8	   & 	0.3 -- 1.1	   \\ 

\midrule
\multicolumn{5}{c}{$e$}\\
\midrule

$\alpha$	  &  deg	  &  0	  &  	 0 -- 32	   & 	 5 -- 56	   \\ 
$d_{\rm 3D,max}$	  &   Mpc	  &  1.0	  &  	1.1 -- 1.9	   & 	1.0 -- 2.4	   \\ 
$TSC_0$	  &   Gyr	  &  0.8	  &  	0.5 -- 0.9	   & 	0.4 -- 1.4	   \\

\bottomrule

\end{tabular}
\end{center}
\end{table}

The MCMAC successfully recovered the gravitationally bound state for each pair with more than 95 per cent of probability. Regarding its accuracy, we were only able to compare $\alpha$, $TTC/TSC_0$ and $d_{\rm 3D,max}$ since the ``Galaxy Cluster Merger Catalog'' does not provide further information about the mock galaxies. Overall, the MCMAC results agree within 68 per cent of confidence level. Same conclusions can be drawn for the execution of MCMAS using no prior constraint in the $v_{\rm plane}$, but the final results have larger error bars. Specifically talking about the time prediction, the MCMAC predictions can be considered in general  as an upper limit of the ``real'' events.

\subsection{A2151}
\label{sec:dis.A2151}

Despite this system have been the subject of many papers, none of them provided with a full description of the entire cluster. We found that the Hercules cluster is comprised of five subclusters: two centrally located and the other three in the periphery. However, the X-ray emission is concentrated on the central part. A2151C-B host a still untouched bright cool core, surrounded by A2151C-F.  The first encounter between them will happen relatively soon, in $0.4$ Gyr. Also, the conjecture of  \cite{Tiwari21} that A2151C-F is a cluster in formation is supported by our analysis.  In this work, we have confirmed quantitatively that A2151S is part of Hercules cluster, being gravitationally bound with it.

We definitely discarded the scenario proposed by \cite{Bird95} in which A2151E and A2152C-F forms a post-merger system, as our analysis has shown that those clusters are not gravitationally bound to each other. In spite of being a plausible scenario, we do not believe that the other subclusters have already experienced a merger, given the absence of a perturbed ICM in the field. Therefore, we suggest that A2151 as a whole is a cluster in an early stage of formation.

\subsection{A2147}
\label{sec:dis.A2147}

We have presented the first description of the internal structure of A2147. This bimodal cluster is the most massive in the field, with a total mass of $\sim13.5 \times 10^{14}$ M$_\odot$ \citep[e.g.][]{Pandge19}. The two subclusters, A2147N and A2147S, are 1.2 Mpc apart from each other (relative to their respective mass/luminosity peaks).

With the background of previous X-ray studies \cite[e.g.][]{Hudson10,Lau12,Lagana19}, there is strong evidence that A2147 as a whole is out of equilibrium. The presence of an offset $\sim$140 kpc \citep{Laine03} between the ICM distribution mapped by its X-ray emission and the BCG \cite[e.g.][]{Kafer19} is commonly referred to as a proxy of a post-merger system. However, this information alone does not allow us to disentangle between the post-merger outgoing or incoming scenarios. We suggest, though, that the outgoing is more likely as the collision happened only 0.8 Gyr ago against 3.0 Gyr for the returning scenario. After such a long time, the observed optical--X-ray offset might vanish.

\subsection{A2152}
\label{sec:dis.A2152}

We have made a significant advance in the comprehension of the internal structure of A2152, despite some questions that have remained still open. One of that is the discrepancy between the mass distribution showing three concentrations whereas the dynamical structure shows only two structures, with a third one overlapped with both A2152N and A2152S1. We conjecture two scenarios: (1) the presence of two concentrations in the projected density map (Fig.~\ref{fig:peaks}) could be an effect of the smooth scale adopted and (2) the inability of the detection of the dynamical counterpart be since the groups have a small separation along the line of sight. Regarding (1), a simple change of the smoothing scale to 11 arcsec can vanish the bimodality and find only one peak midway. About (2), any 3D mixture model will fail to properly classify galaxies in groups with similar redshift.

We have two plausible scenarios for the merging in A2152, both involving a time scale of $\sim$1.4 Gyr (a pre-merger and an outgoing post-merger). To remove this degeneracy, another proxy is necessary. \cite{Blakeslee01} reported an offset between the  BCG and the peak of ICM distribution, but they did not interpret this as a cluster merger signature. Our kinematic analysis of A2152N and A2152S1 pointed that the merger events will happen/happened with a small collision axis, in the sense that a possible detachment between the gas and the visible components (e.g. BCG) would be detectable. So we do not disregard the possibility that the observed offset is real and indicate the A2152 is a post-merger system with its components going to the apoapsis.

\subsection{Hercules supercluster kinematic}
\label{sec:dis.supercluster}

\begin{figure*}
\begin{center}
\includegraphics[width=1.0\textwidth, angle=0]{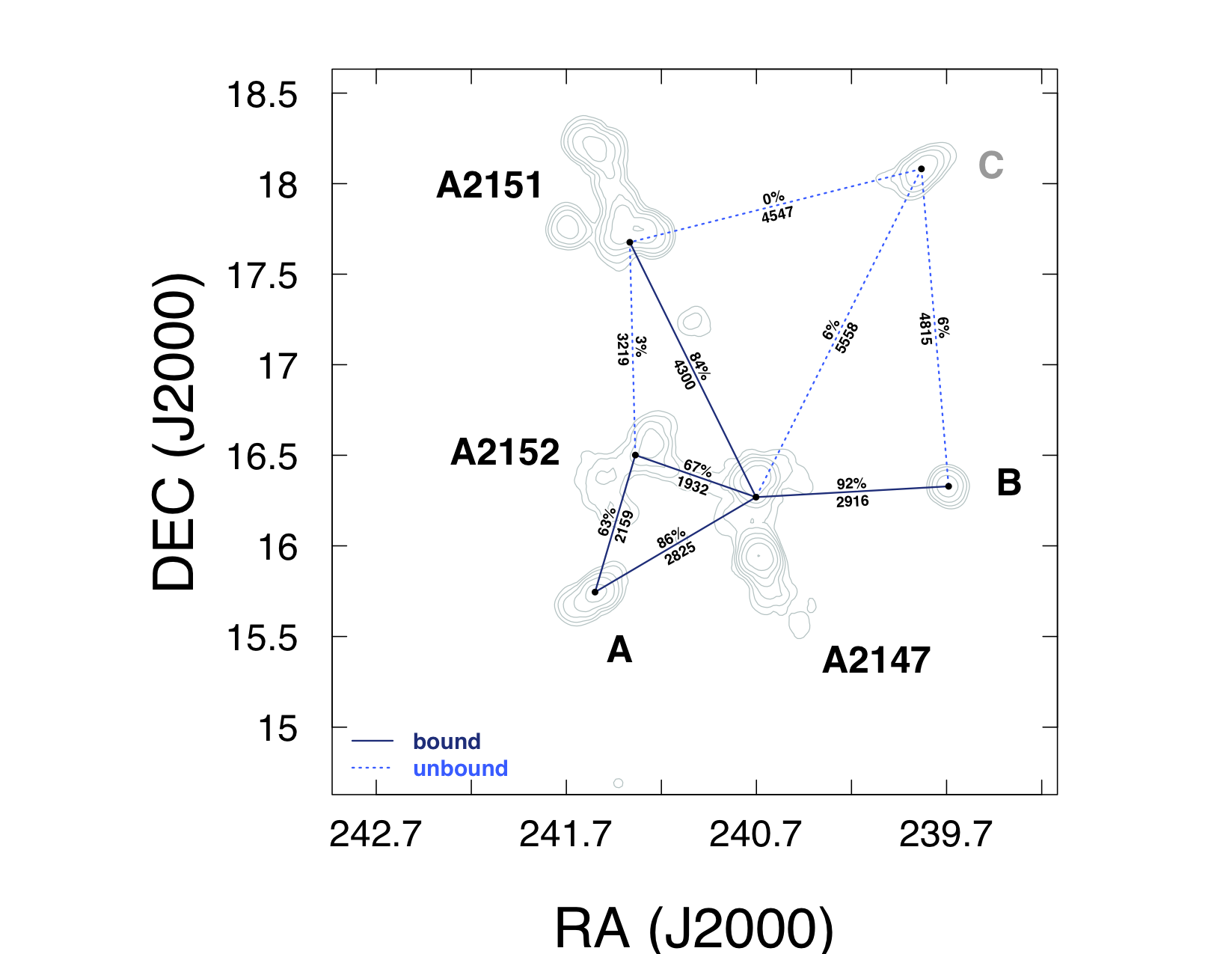} 
\caption{Skeleton of the Hercules supercluster (SCL160) and the dynamic connection among the structures. Besides the previous known members, A2147, A2151, and A2152 we have identified in this work two other, labelled A and B. Straight lines stands for bound pairs and dashed lines for unbound ones.}
\label{fig:Hercules.ske}
\end{center}
\end{figure*}

After the description of each cluster inner structure, we turn our attention to the Hercules supercluster as a whole. To do this, we have considered all clusters as single structures, whose total mass is concentrated at the corresponding mass centre. We also included the three surrounding structures, A, B, and C. and, using the MCMAC-pre, we have investigated the degree of connection in the supercluster. The result is presented in Fig.~\ref{fig:Hercules.ske}.

Our analysis confirms that A2147, A2151, and A2152 are bound to each other. In contrast,  \cite{Barmby98} have found the pair A2147 -- A2152 is not bound.  Even though, due to its projected proximity and the presence of a filament connecting them, we strongly suggest that A2147 -- A2152 are in fact bound, forming the heart of the Hercules supercluster.

Beyond these members, we found other three companion structures in the field. The kinematic analysis has proved that A and B are part of Hercules, as they are gravitationally bound with at least one other member. We have shown that they are single clusters with no signal of substructuring. A search on NED\footnote{NASA/IPAC Extragalactic Database} reported the presence of the galaxy cluster MSPM 00022 at $z=0.03942$ \citep{Smith12} within a projected distance of 8.3 arcmin from A. The same authors report another galaxy cluster, MSPM 00080, at $z=0.03682$ only 2.06 arcmin away from B. We believe that both corresponding to our findings, being this work the first one in including them as Hercules members. On the other side, the structure labelled C is not bound with any other member, therefore not belonging to Hercules supercluster.

As the most massive member, A2147 have the largest number of connections, and it probably will work as a potential well of the future gravitational collapse among all other structures. We estimate the total Hercules mass as $2.1\pm0.2 \times 10^{15}$ M$_\odot$, obtained as a sum of the member's individual mass (Tables \ref{tab:A2151}, \ref{tab:A2147.2152}, and \ref{tab:BC}). This is somewhat lower than proposed by \cite{Barmby98} ($7.6\pm2.0 \times 10^{15}$ M$_\odot$), but unfortunately, the authors do not provide the radius where they computed the masses of the clusters, making a direct comparison not feasible.

The Hercules supercluster emcompases  an approximately  linear region of  $\sim$9 Mpc, being comparatively smaller than other  known systems as A2142 supercluster \citep[50 Mpc;][]{Einasto15},  Ursa Major \citep[50 Mpc;][]{Krause13},  Coma \citep[100 Mpc;][]{Seth20}, and Saraswati \citep[200 Mpc;][]{Bagchi17}, for example. We have shown the presence of a ``bridge'', a filamentary structure, connecting the Northern part of A2147 and A2152. A galaxy concentration is also reported to be located within it (peak \#14 in Fig.\ref{fig:peaks}). This configuration resembles the same as seen in the system A3017/A3016 \citep{Parekh17, Chon19}. Unfortunately, the mixture model was not able to recover its galaxy content, within the criteria we have established to guarantee a high level of confidence in galaxy classification.

We have advanced in the comprehension of the structure and kinematics of the Hercules supercluster, providing an updated description of the system with new ingredients. To finalise, we stress that it consists of a promising target for the mapping of the hot gas along the field, including a filament connecting two clusters. A weak lensing study is also required to provide a detailed map of the mass distribution through the field and determine with more precision, the cluster's masses. Both pieces of information then can be used as input to a tailor-made large scale hydrodynamical simulation in order to describe the details of the supercluster formation.

\section{Summary}
\label{sec:summary}

We summarize the main findings of this work as follows:

\begin{itemize}
    \item The caustic mass, as well those estimated based on the velocity dispersion, is biased through the cluster merger stage, being increased during a short period. This fact is not in line with the caustic principle that the technique is not dependent on the cluster dynamical state. Even though, both techniques provided comparable estimates with our data.
    
    \item A2147 ($\bar{z}=0.0365\pm0.0032$) is a bimodal cluster having a total mass of $13.5_{-1.7}^{+2.1} \times 10^{14}$ M$_\odot$; it is being observed probably after the pericentric passage.
    
    \item A2151 ($\bar{z}=0.0361\pm0.0024$; the Hercules cluster) is composed by five subclusters with a total mass of  $2.88_{-0.27}^{+0.31} \times 10^{14}$ M$_\odot$; it is in an early stage of merger.

    \item A2152 ($\bar{z}=0.0445\pm0.0012$) is comprised by (at least) two subclusters having a total mass of $0.72_{-0.10}^{+0.13} \times 10^{14}$  M$_\odot$.

    \item The core of the Hercules supercluster is constituted by A2147, A2151, and A2152. We found two other galaxy clusters gravitationally bound with this core, increasing the number of known members.
    
    \item The total mass of the Hercules supercluster is estimated in $2.1\pm0.2 \times 10^{15}$ M$_\odot$.

\end{itemize}

\section*{Acknowledgements}
\addcontentsline{toc}{section}{Acknowledgements}

We thank the referee for his/her constructive comments on the work.
VMS acknowledges the CAPES scholarship through the grants 88887.508643/2020-00. ALBR thanks for  the support of CNPq, grant 311932/2017-7. RRdC acknowledges the financial support from FAPESP through the grant \#2014/11156-4.

Funding for the Sloan Digital Sky 
Survey IV has been provided by the 
Alfred P. Sloan Foundation, the U.S. 
Department of Energy Office of 
Science, and the Participating 
Institutions. 

SDSS-IV acknowledges support and 
resources from the Center for High 
Performance Computing  at the 
University of Utah. The SDSS 
website is www.sdss.org.

SDSS-IV is managed by the 
Astrophysical Research Consortium 
for the Participating Institutions 
of the SDSS Collaboration including 
the Brazilian Participation Group, 
the Carnegie Institution for Science, 
Carnegie Mellon University, Center for 
Astrophysics | Harvard \& 
Smithsonian, the Chilean Participation 
Group, the French Participation Group, 
Instituto de Astrof\'isica de 
Canarias, The Johns Hopkins 
University, Kavli Institute for the 
Physics and Mathematics of the 
Universe (IPMU) / University of 
Tokyo, the Korean Participation Group, 
Lawrence Berkeley National Laboratory, 
Leibniz Institut f\"ur Astrophysik 
Potsdam (AIP),  Max-Planck-Institut 
f\"ur Astronomie (MPIA Heidelberg), 
Max-Planck-Institut f\"ur 
Astrophysik (MPA Garching), 
Max-Planck-Institut f\"ur 
Extraterrestrische Physik (MPE), 
National Astronomical Observatories of 
China, New Mexico State University, 
New York University, University of 
Notre Dame, Observat\'ario 
Nacional / MCTI, The Ohio State 
University, Pennsylvania State 
University, Shanghai 
Astronomical Observatory, United 
Kingdom Participation Group, 
Universidad Nacional Aut\'onoma 
de M\'exico, University of Arizona, 
University of Colorado Boulder, 
University of Oxford, University of 
Portsmouth, University of Utah, 
University of Virginia, University 
of Washington, University of 
Wisconsin, Vanderbilt University, 
and Yale University.

The ``Equatorial Red Atlas'' of the southern sky was produced using the UK Schmidt Telescope. Plates from this survey have been digitized and compressedby the STScI. The digitized images are copyright (c) 1992-1995, jointly bythe UK SERC/PPARC (Particle Physics and Astronomy Research Council, formerlyScience and Engineering Research Council) and the Anglo-Australian Telescope Board, and are distributed herein by agreement. All Rights Reserved.

This research has made use of the NASA/IPAC Extragalactic Database, which is funded by the National Aeronautics and Space Administration and operated by the California Institute of Technology.  

This work made use of data from the Galaxy Cluster Merger Catalog (http://gcmc.hub.yt).

\section*{Data availability}
\label{sec:data.ava}

The data underlying this article will be made available under request to the corresponding author.




\bibliographystyle{mnras}
\bibliography{monteiro-oliveira_library}




\bsp	
\label{lastpage}
\end{document}